\documentclass[10pt,preprint]{aastex6}

\usepackage{amsmath,color,xspace}
\newcommand\textlcsc[1]{\textsc{\MakeLowercase{#1}}}
\newcommand{\cii}{[\ion{C}{2}]\xspace}

\shorttitle{Dynamical Characterization of Galaxies at \lowercase{z}$\sim4-6$}
\shortauthors{Jones et al.}

\makeatletter
\DeclareRobustCommand{\atan}{%
  \operatorname{atan}%
  \@ifnextchar2{_}{}%
}
\makeatother

\begin{document}

\pagenumbering{arabic}

\title {Dynamical Characterization of Galaxies at \lowercase{z}$\sim4-6$ via Tilted Ring Fitting to ALMA [CII] Observations}

\author{G. C. Jones\altaffilmark{1,2}, C. L. Carilli\altaffilmark{2,3}, Y. Shao\altaffilmark{4,5}, R. Wang\altaffilmark{4,5}, P. L. Capak\altaffilmark{6,7}, R. Pavesi\altaffilmark{8}, D. A. Riechers\altaffilmark{8}, A. Karim\altaffilmark{9}, M. Neeleman\altaffilmark{10}, F. Walter\altaffilmark{11}} 
\altaffiltext{1}{Physics Department, New Mexico Institute of Mining and Technology, 801 Leroy Pl, Socorro, NM 87801, USA; gcjones@nrao.edu}
\altaffiltext{2}{National Radio Astronomy Observatory, 1003 Lopezville Road, Socorro, NM 87801, USA}
\altaffiltext{3}{Cavendish Astrophysics Group, University of Cambridge, Cambridge, CB3 0HE, UK}
\altaffiltext{4}{Department of Astronomy, School of Physics, Peking University, Beijing 100871, China}
\altaffiltext{5}{Kavli Institute for Astronomy and Astrophysics, Peking University, Beijing 100871, China}
\altaffiltext{6}{Infrared Processing and Analysis Center, California Institute of Technology, MC 100-22, 770 South Wilson Ave., Pasadena, CA 91125, USA}
\altaffiltext{7}{Spitzer Science Center, California Institute of Technology, Pasadena, CA 91125, USA}
\altaffiltext{8}{Department of Astronomy, Cornell University, Space Sciences Building, Ithaca, NY 14853, USA}
\altaffiltext{9}{Argelander-Institut f\"{u}r Astronomie, Universit\"{a}t Bonn, Auf dem H\"{u}gel 71, D-53121 Bonn, Germany}
\altaffiltext{10}{University of California Observatories-Lick Observatory, University of California, Santa Cruz, CA, 95064, USA}
\altaffiltext{11}{Max-Planck-Institut f\"{u}r Astronomie, K\"{o}nigstuhl 17, D-69117 Heidelberg, Germany}


\begin{abstract}

Until recently, determining the rotational properties of galaxies in the early universe ($z>4$, Universe age $<1.5$\,Gyr) was impractical, with the exception of a few strongly lensed systems. Combining the high resolution and sensitivity of ALMA at (sub-) millimeter wavelengths with the typically high strength of the \cii 158\,$\mu$m emission line from galaxies and long-developed dynamical modeling tools raises the possibility of characterizing the gas dynamics in both extreme starburst galaxies and normal star forming disk galaxies at $z\sim4-7$. Using a procedure centered around GIPSY's \textlcsc{ROTCUR} task, we have fit tilted ring models to some of the best available ALMA \cii data of a small set of galaxies: the MS galaxies HZ9 \& HZ10, the Damped Lyman-alpha Absorber (DLA) host galaxy ALMA J0817+1351, the submm galaxies AzTEC/C159 and COSMOS J1000+0234, and the quasar host galaxy ULAS J1319+0950. This procedure directly derives rotation curves and dynamical masses as functions of radius for each object. In one case, we present evidence for a dark matter halo of $\mathcal{O}(10^{11}\,)$M$_{\odot}$. We present an analysis of the possible velocity dispersions of AzTEC/C159 and ULAS J1319+0950  based on matching simulated observations to the integrated \cii line profiles. Finally, we test the effects of observation resolution and sensitivity on our results. While the conclusions remain limited at the resolution and signal-to-noise ratios of these observations, the results demonstrate the viability of the modeling tools at high redshift, and the exciting potential for detailed dynamical analysis of the earliest galaxies, as ALMA achieves full observational capabilities.

\end{abstract}


\section{Introduction}

Rotation curves of galaxies reveal the underlying mass structure, hint
at past mergers or gravitational interactions, and allow for simple
dynamical mass estimates \citep{sofu01}. These profiles of velocity as a function of radius have been
regularly measured for local galaxies since \citet{peas18} deduced the rotation
curve of M31. Subsequent observations of this source (\textit{e.g.}, \citealt{babc39,rubi70})
showed an initial rise in velocity near the center of the galaxy
followed by a nearly constant velocity out to large radius. This
profile type was confirmed in other galaxies (\textit{e.g.},
\citealt{burb60,bosm78}), leading to the question of its cause.
Solid-body rotation would give a linear relationship between velocity
and radius, while Keplerian rotation would result in decreasing velocity
at large radius. The sum of the masses of the baryonic components as a
function of radius was shown to be insufficient to cause the flat
rotation curves, necessitating a ``dark matter'' component to give rise
to the flat rotation curves.

The question of galaxy dynamics in general, and of the dark matter
content specifically, of disk galaxies at high redshift is only
recently coming into consideration with the advent of new large
facilities. The dark matter content of early-type galaxies at high
redshift has been inferred from gravitational lensing (\textit{e.g.}, 
\citealt{lang90}). For late-type galaxies, some progress on determining
dynamics at $z \sim 1-2$, has been made using integral field units
and the H$\alpha$ line (\textit{e.g.} \citealt{wuyt16}). Wuyts et al. find
evidence for rotating disks, and that baryons comprise most ($> 50\%$)
of the dynamical mass within the optical disks of galaxies out to $z
\sim 2$. A few cases of galaxy dynamics based on CO observations at $z
\sim 1-4$ have been published, but in all cases the galaxies are
the most extreme massive starbursts, and even then, the resolution and
signal to noise has been limited (\citealt{riec08,tacc10,hodg12,tacc13}; see \citealt{cariw13} for a
review).

Tilted ring models have been used as a standard analysis tool to
characterize the rotation of galaxies since \citet{rogs74}. They
consist of a number of rings with variable width, systemic velocity,
position angle, inclination, and rotational velocity. By varying each,
creating a model velocity field, and comparing it to the observed
velocity field, the dynamical nature of the object can be
determined. While they allow for the existence of warps (\textit{i.e.}, 
variations in inclination angle across the disk),
counter-rotation, and velocity gradients, their azimuthal averaging
wipes out any inherent information regarding spiral arm structure or
central bars.

However, the fitting process directly outputs the rotation
curve, allowing for dynamical mass calculation and kinematic
characterization. These dynamical masses may then be used to test the relationship 
between the mass of a galaxy and the mass of its central black hole (see \citealt{korm13} for a review)
or to probe the amount of dark matter present (\textit{e.g.}, \citealt{rich15}).
In addition, a detection of the flat portion of a
rotation curve argues that the underlying structure is a rotating
disk, not a velocity gradient due to a merger event or infall/outflow
(\textit{e.g.}, \citealt{garc14,ueda14}). We employ the tilted ring modeling
herein, to create rotation curves and radial dynamical mass profiles of $z\sim4-6$ galaxies.

Excellent rotation curves have been created for local objects using \ion{H}{1} 
data (\textit{e.g.},
\citealt{vogt96,debl08,schi14,rich15,rich16}). These observations have the benefits
of high brightness and large spatial extent. However, this tracer does
not reveal the dynamics of the central region of the disk, due to the low density of \ion{H}{1} in the center of galaxies \citep{wang14}. An
alternative is H$\alpha$, which traces ionized gas and has been used
to observe the kinematics of both $z\sim1-3$ galaxies (\textit{e.g.},
\citealt{genz06,vans08,wrig07,epin12,burk16}) and local galaxies
(\textit{e.g.}, \citealt{pers95}). Additionally, CO (\citealt{fran16}), which
traces molecular gas, and [\ion{C}{2}], which traces both \ion{H}{2} regions
and colder gas (\citealt{sofu01,vene16}), may be used.

The 158$\mu$m line of [\ion{C}{2}] ($^2P_{3/2}\rightarrow^2P_{1/2}$) has two
pointed advantages in dynamical analysis of early galaxies. First is
that the line is typically the brightest emission line at wavelengths
from the radio through the FIR from star forming galaxies. And second,
[\ion{C}{2}] emission comes from multiple phases of the ISM, including
the warm ionized medium, warm and cold atomic medium, and warm dense
molecular medium \citep{pine13}. While this generally complicates
interpretation of [\ion{C}{2}] luminosity, this wide range of physical
conditions makes [\ion{C}{2}] an excellent tracer of kinematics over large
areas of the disk. It is also excited by a range of sources, including collisions 
with electrons, hydrogen atoms, and H$_2$ \citep{gold12}. It can act as a tracer of molecular hydrogen in ``dark'' 
clouds deficient in CO \citep{lang10}. 

It should be noted that \cii traces a smaller region than other dynamical tracers (\textit{i.e.}, HI, H$\alpha$, CO; 
\citealt{debl16}). Because of this, the outer disk will likely not be observable. While its rest-frame frequency of
1.9\,THz (158\,$\mu$m) requires space-based telescopes (or the Stratospheric
Observatory for Infrared Astronomy; SOFIA) for study of nearby
galaxies, emission from high-$z$ objects is redshifted into the ALMA ($z>1$) and NOEMA ($z>6$)
bands. With the advent of ALMA and the use of redshifted [\ion{C}{2}], it is
possible to extend classical dynamical analysis to the first galaxies,
possibly revealing the dark matter content therein. 

Here, we present the results of using long-developed tools to fit tilted ring
models to a sample of two Lyman break galaxies (LBGs), a damped
Lyman-$\alpha$ absorber (DLA) host galaxy, two submillimetre
galaxies (SMGs), and a QSO host galaxy, all between $z=4.2-6.2$, as
observed with ALMA in [\ion{C}{2}]. In addition, we use a limited form of three-dimensional dynamical fitting
to constrain the velocity dispersions and density profiles of one of the above SMGs and the QSO host galaxy.
The paper is organized as follows: 
we detail the observations of the sources in our sample in Section \ref{sources};
the procedure to fit tilted rings to these sources is explained in Section \ref{method1},
with the results of this process in Section \ref{results};
an alternate fitting procedure is discussed in Section \ref{method2};
the accuracy of our derived parameters is discussed in Section \ref{PEA};
and we summarize in Section \ref{conclude}. 

We will assume ($\Omega_{\Lambda}$,$\Omega_m$,h)=(0.682,0.308,0.678) \citep{plan16} throughout.

\section{Sources} \label{sources}
Our sample consists of six galaxies at $z=4\sim6$ observed in [\ion{C}{2}] with ALMA. 
Selection was based on the presence of a velocity gradient that was well modeled by a 
tilted ring model. 

Multiple other sources were considered, but not included in our sample, including HZ6 in the \citet{capa15} sample, the SMG in BRI1202--0725 ($z\sim4.5$; \citealt{cari13}), the SMGs Vd-17871 ($z\sim4.6$; A. Karim et al., in preparation), AzTEC-3 ($z\sim5.3$; \citealt{riec14}), the QSO host galaxy ULASJ2315+0143 ($z\sim2.6$; \citealt{bane17}), and the local main sequence galaxy NGC4191 (L. Young et al., in preparation). 
HZ6 has the third highest \cii detection in the same sample as two galaxies that were successfully fit, but our spectral resolution was too low to allow for successful fitting. Similarly, our spatial resolution for the SMG in 
BRI1202--0725 was too low. The SMGs Vd-17871 and AzTEC-3 showed gradients that were too weak to fit. Both ULASJ2315+0143 and NGC4191 were well fit, but were published separately due to their line emission being CO rather than [\ion{C}{2}].

It may seem strange that the sample of galaxies that we successfully fit (main sequence galaxies, SMGs, and a QSO host galaxy) are qualitatively similar to those that we failed to fit (SMGs and a QSO host galaxy). This shows that while our procedure is applicable to a range of galaxy types, the intrinsic velocity gradient must be strong enough to be observed at a given spatial and spectral resolution. In addition, the rotation must be orderly and disk-like (\textit{i.e.}, not early-stage major mergers or outflow-dominated galaxies).

Below, we detail the observations of each source that was successfully fit.

\subsection{HZ9 and HZ10}
HZ9 ($z=5.5410$) and HZ10 ($z=5.6566$) were originally selected as LBGs in the COSMOS field \citep{leau07}, and were observed in [\ion{C}{2}] and rest-frame-FIR emission in cycle 1 ALMA by \citet{capa15}. Both HZ9 and HZ10 were detected in rest-frame $150\,\mu$m emission ($0.52\pm0.04$\,mJy and $1.26\pm0.04$\,mJy) and [\ion{C}{2}] ($1.95\pm0.07$\,Jy km s$^{-1}$ and $4.5\pm0.3$\,Jy km s$^{-1}$). The star formation rate (SFR) of each source was estimated using SED fits to greybody models based on the $158\,\mu$m continuum flux density and source intensities in COSMOS and Ultra-Vista images. A range of greybody parameters was assumed from observations of $z>2$ objects, resulting in SFRs of ($70\pm30$\,M$_{\odot}$ year$^{-1}$ and $170\pm30$\,M$_{\odot}$ year$^{-1}$). [\ion{N}{2}] emission from HZ10 was detected by \citet{pave16}.

ALMA [\ion{C}{2}] observations for HZ9/HZ10 (project 2012.1.00523.S) used 27/25 antennas, consisted of 40/50 minutes of on-source time, and were undertaken on November 14/16, 2013. These band 7 observations resulted in spatial resolutions of $\sim0.6''$ and a spectral resolution of $\sim15$\,km s$^{-1}$ for the [\ion{C}{2}] line. Our HZ9 data were taken from directly from \citet{capa15}. For HZ10, we use the updated calibration by \citet{pave16}.

\subsection{J0817+1351}
ALMA J081740.85+135138.2 (J0817+1351) is the host galaxy of a high metallicity damped Lyman-alpha absorber (DLA) at $z = 4.2584$. The DLA at $z =4.2584$ was originally detected in Ly-$\alpha$ absorption towards the background quasar at a redshift of $z = 4.398$. These observations yielded an \ion{H}{1} column density of $\log (N$(\ion{H}{1})/cm$^{-2}$) $= 21.3 \pm 0.2$ and a metallicity of [M/H] $= -1.15 \pm 0.15$ \citep{rafe12}. 

The [\ion{C}{2}] 158\,$\mu$m observations come from \citet{neel17}, who observed J0817+1351 on December 30, 2015 with ALMA in cycle 3 (project 2015.1.01564.S). The total on-source time was 46 minutes with a spatial resolution of $\sim 1''$ and spectral resolution after smoothing of $\sim 50~{\rm km~s}^{-1}$. [\ion{C}{2}] emission was found $\sim6''$ from the quasar at the redshift of the DLA. In addition to the detection of [\ion{C}{2}], the detected continuum emission at 158\,$\mu$m implies a SFR of $110 \pm 10$\,M$_\odot$\,year$^{-1}$. The maximum rotational velocity was determined from fitting an exponential disk to the data (see \citealt{neel16}), resulting in a dynamical mass limit of $> 6 \times 10^{10}$~M$_\odot$.

\subsection{AzTEC/C159}
Located in the COSMOS field, AzTEC/C159 ($z=4.5670$) is an SMG discovered at 1.1\,mm in the AzTEC/ASTE survey \citep{arex11}. \citet{smol15} derived a dust temperature of $T_{\rm dust}=39\pm2$\,K, SFR$\sim700\pm200$\,M$_{\odot}$ year$^{-1}$, and $M_{*}=1.1\times10^{11}$\,M$_{\odot}$. Smol{\v c}i{\'c} et al. determined the dust mass using three techniques: fitting the NUV-mm SED with MAGPHYS ($M_{\rm dust}=1.6\times10^9$\,M$_{\odot}$), fitting the FIR SED assuming an optically thin blackbody ($M_{\rm dust}=(5.0^{+1.3}_{-1.0})\times10^8$\,M$_{\odot}$), and fitting the FIR SED assuming the dust model of \citet{drai07} ($M_{\rm dust}=(2.0^{+3.0}_{-1.2})\times10^9$\,M$_{\odot}$). We will use the last dust mass in the following analysis.

Additional observations were made (E. Jimenez-Andrade et al., in preparation) using the VLA in D- and DnC-configurations and NOEMA in D- and C-configurations. These detected emission in CO(5-4) and CO(2-1), resulting in a molecular gas mass $M_{\rm H_2}=(1.1\pm0.3)\times10^{10}$\,M$_{\odot}$ (assuming $\alpha_{CO}=0.8\,M_{\odot}\,$K$^{-1}$\,km$^{-1}$\,s\,pc$^{-2}$, appropriate for ULIRGs; \citealt{solo05}).

ALMA [\ion{C}{2}] observations for AzTEC/C159 (project 2012.1.00978.S; A. Karim et al., in preparation) used 34 antennas, consisted of 20 minutes of on-source time, and were undertaken on June 15, 2014. These band 7 observations resulted in spatial resolutions of $\sim0.4''$ and a spectral resolution of $\sim16$\,MHz$\sim14$\,km s$^{-1}$ for the [\ion{C}{2}] line.

\subsection{J1000+0234}
COSMOS J100054+02343 (J1000+0234; $z=4.542$) is a galaxy in the COSMOS field. Originally detected as a V-band dropout \citep{cari08}, followup observations and SED fitting (visible to cm) revealed a possible major merger with a SFR$>10^3$\,M$_{\odot}$ year$^{-1}$ and $L_{\rm IR}\sim10^{13}$\,L$_{\odot}$ \citep{capa08}. Observations of CO(4-3) and CO(2-1) suggest a gas mass of $M_{\rm H_2}$=$2.6\times10^{10}$\,M$_{\odot}$ (assuming $\alpha_{CO}=0.8\,M_{\odot}\,$K$^{-1}$\,km$^{-1}$\,s\,pc$^{-2}$), and a dynamical mass of $1.1\times10^{11}$M$_{\odot}$ \citep{schi08}.

ALMA [\ion{C}{2}] observations for J1000+0234 (project 2012.1.00978.S; A. Karim et al., in preparation) used 39 and 36 antennas on June 2 \& June 14, 2014, resulting in a combined on source time of 44 minutes. These band 7 observations resulted in spatial resolutions of $\sim0.3''$ and a spectral resolution of $\sim16$\,MHz$\sim14$\,km s$^{-1}$ for the [\ion{C}{2}] line.

\subsection{J1319+0950}
ULAS J131911.29+095051.4 (J1319+0950;$z=6.127$)is a QSO detected in the UKIRT Infrared Deep Sky Survey (UKIDSS) \citep{mort09}. By fitting an optically thin greybody model with T$_{\rm dust}=47$\,K and $\beta=1.6$ to their FIR SED, \citet{wang11} estimate $L_{\rm FIR}$=$1.0\times10^{13}$\,L$_{\odot}$. Using this FIR luminosity and an altered form of equation 1 of \citet{bian13}, they find a dust mass of $5.7\times10^8$\,M$_{\odot}$.

It has been previously observed with cycle 0 ALMA (project 2011.0.00206.S) in [\ion{C}{2}] \citep{wang13}, where a velocity gradient was observed. They estimate the dynamical mass to be $M_{\rm dyn}=1.2\times10^{11}$M$_{\odot}$ at an inclination angle of $56^{\circ}$ (from the ratio of the \cii major and minor axes), using
\begin{equation} \label{fwhm}
M_{\rm dyn}\propto \left( FWHM_{\rm [CII]}/sin(i) \right)^2 D
\end{equation}
where $FWHM_{\rm [CII]}$ is the full width at half maximum of the \cii line, $i$ is the inclination angle, and $D$ is the radius of the galaxy. Note that this is simply assuming that the \cii - emitting material is rotating in a purely circular fashion, and that we will make this assumption as well.

More recently, ALMA [\ion{C}{2}] observations for J1319+0950 (project 2012.1.00240.S; \citealt{shao17}) used 34 antennas in August 2014, resulting in a combined on source time of 36 minutes. These band 7 observations resulted in spatial resolutions of $\sim0.2''$ and a spectral resolution of $62.5$\,MHz$\sim70$\,km s$^{-1}$ for the [\ion{C}{2}] line.

The rotation of J1319+0950 has been fitted by \citet{shao17}, using a similar method as adopted here. We will compare their results with ours, along with further analysis of the density profile.

\section{Tilted Ring Fitting Methods} \label{method1}
We began with ALMA [\ion{C}{2}] image cubes of each source. These cubes were converted from frequency to velocity space by assuming zero velocity at the redshifted frequency of the [\ion{C}{2}] line ($\nu_{rest}=$1900.5369\,GHz), where redshifts were taken from previously published [\ion{C}{2}] observations. Velocity fields were made from these cubes and run through the GIPSY task \textlcsc{rotcur} \citep{vand92} to find the rotation curve of the source. This rotation profile was used in GIPSY's \textlcsc{velfi} to create a model velocity field, which we subtracted from a velocity field created from the data to test the goodness of fit. 

\subsection{Velocity Fields}
As discussed by \citet{debl08}, multiple methods exist for determining the velocity behavior of a source. The most common is the first moment or velocity weighted image, implemented in CASA as \textlcsc{immoments}:
\begin{equation}
M_1=\int{I_vvdv}\bigg/\int{I_vdv}
\end{equation}
This is well suited for high S/N systems with strong lines, as the entire line profile contributes to each integral. However, if the line is weak, noise contributions may render any velocity information unreachable.

The other main methods examine the spectrum of each spatial pixel in an image cube. The velocity may be found from the velocity with the maximum amplitude (\textit{i.e.}, peak velocity field) or from the centroid velocity of a Gaussian fit to the spectral line. While these two approaches are identical in the case of strong lines, the latter is more appropriate for weaker systems. The fit may be be a single or double component fit, or may include an asymmetric term. 

Due to the faintness of high redshift sources and the limited (albeit considerable) sensitivity of ALMA, our sources have relatively low S/N, rendering first moment maps useful only as rough indicators of velocity gradients. The best velocity fields were found using the single Gaussian fitter in the AIPS task \textlcsc{xgaus}. This interactive fitter allowed for rejection of poor fits, resulting in reliable velocity fields.

\subsection{Ring Widths}
Before we could begin fitting tilted ring models to these velocity fields, the possible dimensions of these rings had to be quantified. Due to the small angular sizes of these sources, we assumed that they could be approximated as flat, warp-less (i.e., constant inclination angle) disks. The simplest model would be a single ring spanning the entire velocity field, but this does not satisfy Nyquist sampling. Instead, we adopt two limits on ring width: an upper limit of the full width at half maximum (FWHM) of the minor axis of the synthesized beam divided by three and a lower limit of the FWHM of the minor axis of the synthesized beam divided by two. The resulting number of possible rings ($N_{\rm Rings}$) and the maximum radial extent of each galaxy ($R_{\rm max}$) are listed in Table \ref{ringwidth}.

\begin{deluxetable}{lccc}
\tablecolumns{4}
\tablewidth{0pt}
\tablecaption{Ring Width Limits \label{ringwidth}}
\tablehead{ \colhead{Source} & \colhead{$R_{\rm max}$[$''$]} & \colhead{$N_{\rm Rings,MIN}$} & \colhead{$N_{\rm Rings,MAX}$}}
\startdata
HZ9 & 0.69 & $2$ & 4\\ 
HZ10 & 0.85 & $3$ & 5\\
J0817+1351 & 1.20 & $2$ & 4\\
AzTEC/C159 & 0.53 & $2$ & 4\\
J1000+0234 & 0.70 & $5$ & 7\\
J1319+0950 & 0.58 & $6$ & 8
\enddata
\end{deluxetable}

Previous methods (\textit{e.g.}, \citealt{debl08,rich15}) used a single number of rings to create rotation curves. Since we assume a warp-less disk, we consider the results of combinations of ring numbers. These ring limits are influenced by the range in Table \ref{ringwidth} (\textit{i.e.}, $N_{\rm Rings,MIN}$ to $N_{\rm Rings,MAX}$), but are varied to include all possible permutations (\textit{e.g.}, 2, 3, 4, 2-3, 3-4, 2-4). In this way, we consider six sets of ring numbers for each source.

\subsection{Rotation Fitting}\label{RF}
\begin{figure}[h]
\centering
\includegraphics[scale=0.8,clip=true]{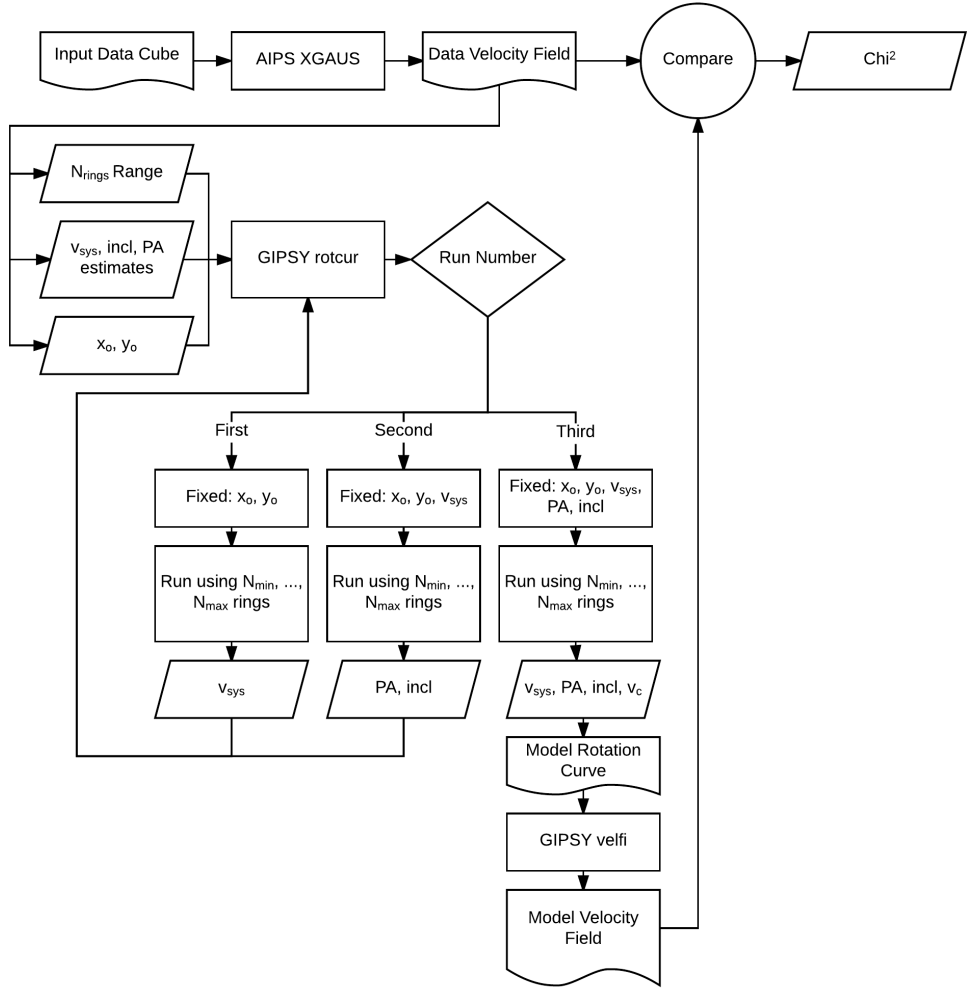}
\caption{Outline of iterative fitting process}
\label{flow}
\end{figure}

The foundations of the approach described in this section are based on the methods of \citet{debl08} and \citet{rich15}, as outlined in Figure \ref{flow}. They have been applied to derive rotation curves for The \ion{H}{1} Nearby Galaxy Survey (THINGS; \citealt{debl08}) and as a step in the dark matter mass decomposition of NGC5005 \citep{rich15}. Both this work and those upon which it is based assume purely circular rotation (\textit{i.e.}, $v_{\rm radial}=0$), which is considered a valid approximation for well-ordered disks. However, due to our low S/N sources, we will assume a constant inclination and position angle across each disk, rather than allowing for a light variation in each.

The velocity fields from \textlcsc{xgaus} were fed into the GIPSY task \textlcsc{rotcur}, which fit a series of tilted rings to the source. Each ring had a set radius ($R$) and width, but variable rotation velocity ($v_{\rm c}$), inclination angle ($i$), position angle ($\phi$), center ($x_{\rm o}$,$y_{\rm o}$), systemic velocity ($v_{\rm sys}$), and expansion velocity (which we assumed to be zero). These are related to the observed line of sight velocity $V(x,y)$ by
\begin{equation} 
V(x,y)=v_{\rm sys}+v_{\rm c}(r)\sin(i)\cos(\theta)
\label{eq1}
\end{equation}
where $\theta$ is the position angle of a point (x,y) in the plane of the observed galaxy (\textit{i.e.}, not in the plane of the sky) with respect to the receding semimajor axis \citep{debl08}:
\begin{equation}
\cos(\theta)=\frac{-(x-x_{\rm o})\sin(\phi)+(y-y_{\rm o})\cos(\phi)}{R}
\label{eq2}
\end{equation}
By direct comparison of the input velocity field and the velocity field created by the model rings, \textlcsc{rotcur} fits for the best ring parameters.

The results are slightly dependent on the initial estimates of each parameter (central position, $v_{\rm sys}$, $\phi$, $i$). To counteract this, we used two methods of deriving estimates. The first was to obtain estimates of the position angle, inclination, and systemic velocity from the input velocity field by eye, then to run \textlcsc{rotcur} to verify their plausibility (RC1). The second method found the systemic velocity by fitting a 1-D Gaussian to the [\ion{C}{2}] line profile and the other parameters by fitting a 2-D Gaussian to the [\ion{C}{2}] zeroth moment image (GAUS). When deriving our final curves, we considered all combinations of parameters from each method.

In what follows, a `run' begins with an estimate of central position, $v_{\rm sys}$, $\phi$, and $i$, along with a set range of ring widths, and ends with a derived rotation curve and a set of best-fit parameters.

We begin by fixing the ring centers to be identically equal to an estimate (RC1 or GAUS), and setting a minimum and maximum number of rings. Letting all other variables change, we run \textlcsc{rotcur} for all current permutations of ring number, starting with a run using $N_{\rm min}$ rings and then repeating with a run using $N_{\rm min}+1$ rings, and so on until a run using $N_{\rm max}$ rings. 

The run results are aggregated, and unsuccessful rings are discarded. Success is defined as: $v_{\rm c}>3\times\delta v_{\rm c}$, $\delta i<50^{\circ}$, and $\delta \phi<50^{\circ}$, where $\delta$ signifies uncertainty. An average of the systemic velocities of all successful rings is taken, weighted by their fitting uncertainties. Fixing $v_{\rm sys}$ to be this averaged value, we then repeat the fitting procedure, but fitting for $\phi$ and $i$ simultaneously. We then fix $\phi$ and $i$ to their derived averages.

In the final run, the only variable is $v_{\rm c}$. The set of successful rings with the smallest width are chosen as representative of the overall rotation. The uncertainty in rotational velocity for each ring is estimated as the maximum of the uncertainty by \textlcsc{rotcur} for that ring, the velocity width of a channel, and $v_c|sin(i)-sin(i\pm\delta\,i)|$. In this way, we include the spectral resolution of the input cube, the goodness of fit from \textlcsc{rotcur}, and the effects of inclination uncertainty. Another common estimate of the uncertainty is the difference in rotational velocities for the receding and approaching halves of the galaxy \citep{debl08}, but we lack the signal to perform this.

In order to test the goodness of fit, we use GIPSY's \textlcsc{velfi} task to create a velocity field based on the rotation curve, $\phi$, $i$, and $v_{\rm sys}$ that were found from \textlcsc{rotcur}. This field, which is created using the same spatial resolution as our data, is then compared to the field of the original data. The residual velocity values for each pixel are added in quadrature and normalized by the number of pixels, resulting in a normalized error value. 

As a summary example, consider HZ9, which had $N_{\rm Rings,MAX}=4$ and $N_{\rm Rings,MIN}=2$. By changing the initial estimates for each variable (center position, $v_{\rm sys}$, $\phi$, $i$) between those of our two methods (\textit{i.e.}, by eye and by fits to moment images), we have $2^4=16$ possible combinations of initial estimates. There are three runs using a single number of rings (2, 3, 4), two runs using a range of two rings (2-3, 3-4), and one run using a range of three rings (2-4), resulting in six possible combinations. By varying both parameter estimates and ring number, we thus consider the results of $16\times6=96$ full runs. Similarly, we considered 96 full runs for HZ10, J0817+1351, AzTEC/C159, and J1000+0234.

\section{Results \& Analysis} \label{results}
All six objects were well fit by the tilted ring models, with their resulting rotation curves varying in detail. Table \ref{fitres} shows the resulting redshifts (found from $v_{\rm sys}$), position angles, inclinations, and positions of the best model for each source. For J1319+0950, both our results and those of \citet{shao17} are given. 

For each source, the results are presented from our initial \textlcsc{rotcur} approach (RC1), fitting 1-D Gaussians to the line profile and 2-D Gaussians to the zeroth moment images (GAUS), and our full exploration of the parameter space (RC2). Note that the position angles are measured counter-clockwise from north, and numbers in parentheses represents uncertainty in the last digit. As a conservative estimate of the uncertainty in redshift, we took the greater of the implied uncertainty from $v_{\rm sys}$ from \textlcsc{rotcur} and the velocity width of a channel in the initial image cube. 

It should be noted that the uncertainties in Table \ref{fitres} for the position angle and inclination are those outputted by \textlcsc{rotcur}, not including uncertainties from errors from observational effects, and thus are underestimated. See the Appendix for a discussion of this error. Uncertainties in positions are $5\times$ the cell sizes of the original image cubes, while uncertainties in $R_{\rm max}$ are half of the ring width.
\pagebreak

\begin{deluxetable}{lc|ccccccc}
\tablecolumns{9}
\tablewidth{0pt}
\tablecaption{Rotcur Fitting Results \label{fitres}}
\tablehead{ \colhead{Source} & & \colhead{$z_{\rm fit}$} & \colhead{$\phi$ [$^{\circ}$]} & \colhead{$i$ [$^{\circ}$]} & \colhead{RA} & \colhead{Dec} & \colhead{$R_{\rm max}$ [kpc]} & \colhead{$v_{\rm max}$ [km s$^{-1}$]}}
\startdata
HZ9   & RC1  & $5.5415(1)$ & $11\pm4$   & $60\pm8$  & 9h59m51.70(4)s  & +02$^{\circ}$22$'$42.1(5)$''$&&\\ 
      & GAUS & $5.5410(1)$ & $84\pm11$  & $45\pm9$  & 9h59m51.69(4)s  & +02$^{\circ}$22$'$42.1(5)$''$&&\\
      & RC2  & $5.5417(3)$ & $5\pm6$    & $51\pm9$  & 9h59m51.69(4)s  & +02$^{\circ}$22$'$42.1(5)$''$& $3.8\pm0.6$ & $196\pm16$\\[2mm]
HZ10  & RC1  & $5.6533(3)$ & $294\pm5$  & $62\pm12$ & 10h0m59.31(4)s  & +01$^{\circ}$33$'$19.4(5)$''$&&\\
	  & GAUS & $5.6541(5)$ & $273\pm4$  & $70\pm5$  & 10h0m59.32(4)s  & +01$^{\circ}$33$'$19.4(5)$''$&&\\
	  & RC2  & $5.6532(4)$ & $299\pm1$  & $56\pm4$  & 10h0m59.31(4)s  & +01$^{\circ}$33$'$19.4(5)$''$ & $4.2\pm0.6$ & $367\pm44$\\[2mm]
J0817+1351   & RC1  & $4.2602(1)$  & $89\pm9$   & $44\pm9$  & 8h17m40.85(4)s  & +13$^{\circ}$51$'$38.3(5)$''$&&\\
      & GAUS & $4.2604(4)$  & $62\pm15$  & $60\pm23$ & 8h17m40.85(4)s  & +13$^{\circ}$51$'$38.2(5)$''$&&\\
      & RC2  & $4.2605(9)$ & $98\pm1$   & $43\pm8$  & 8h17m40.85(4)s  & +13$^{\circ}$51$'$38.2(5)$''$& $7\pm1$ & $228\pm52$\\[2mm]
AzTEC/C159 & RC1  & $4.5666(8)$ & $174\pm3$  & $33\pm11$ & 9h59m30.42(2)s  & +01$^{\circ}$55$'$27.5(3)$''$&&\\
      & GAUS & $4.5664(3)$ & $207\pm45$ & $31\pm21$ & 9h59m30.42(2)s  & +01$^{\circ}$55$'$27.6(3)$''$&&\\
      & RC2  & $4.5665(3)$ & $174\pm1$  & $28\pm6$ & 9h59m30.42(2)s  & +01$^{\circ}$55$'$27.5(3)$''$& $4.1\pm0.6$ & $718\pm18$\\[2mm]
J1000+0234 & RC1  & $4.540(1)$  & $148\pm2$ & $59\pm7$  & 10h0m54.49(2)s  & +$2^{\circ}34'36.1(2)''$&&\\
           & GAUS & $4.539(1)$  & $147\pm4$ & $66\pm3$  & 10h0m54.48(2)s  & +$2^{\circ}34'36.1(2)''$&&\\
           & RC2  & $4.540(1)$  & $148\pm2$ & $53\pm7$ & 10h0m54.49(2)s  & +$2^{\circ}34'36.1(2)''$& $4.4\pm0.3$ & $635\pm25$ \\[2mm]
J1319+0950 & RC1       & $6.133(2)$  & $57\pm5$ & $24\pm14$  & 13h19m11.29(2)s & +$09^{\circ}50'51.6(2)''$&&\\
           & GAUS      & $6.134(2)$  & $45\pm13$ & $50\pm10$ & 13h19m11.29(2)s & +$09^{\circ}50'51.4(2)''$&&\\
           & RC2       & $6.133(2)$  & $56\pm10$ & $29\pm3$ & 13h19m11.29(2)s & +$09^{\circ}50'51.6(2)''$& $3.1\pm0.3$ & $465\pm70$ \\
           & SHAO$^a$  & $6.133(2)$  & $56\pm4$  & $33\pm3$ & 13h19m11.29(2)s & +$09^{\circ}50'51.6(2)''$& $3.2\pm0.3$ & $427\pm70$
\enddata
\tablecomments{Errors may be underestimated (see Appendix). \textit{a}-See \citet{shao17} for details of fitting procedure.}
\end{deluxetable}

\begin{figure}[b]
\centering
\includegraphics[scale=0.4,clip=true]{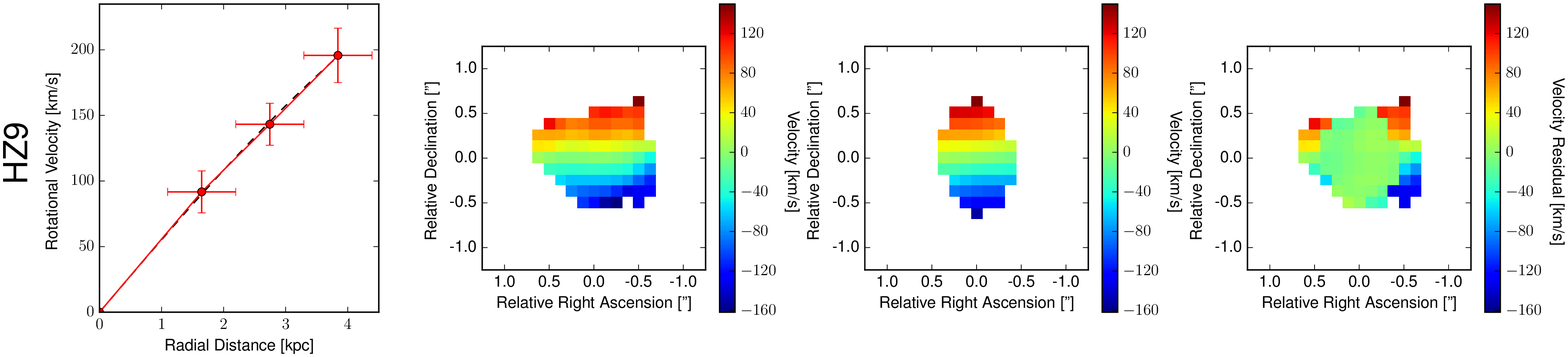}
\includegraphics[scale=0.4,clip=true]{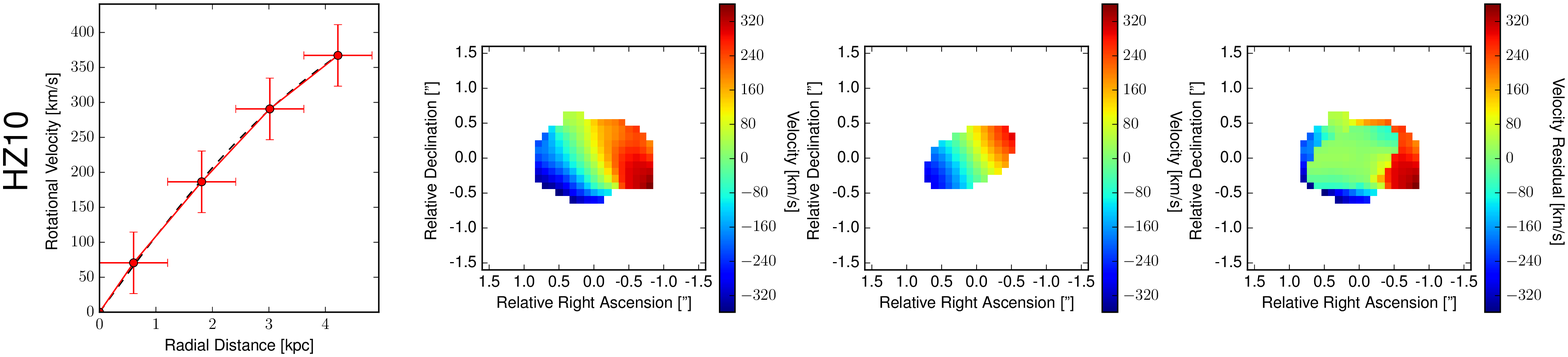}
\includegraphics[scale=0.4,clip=true]{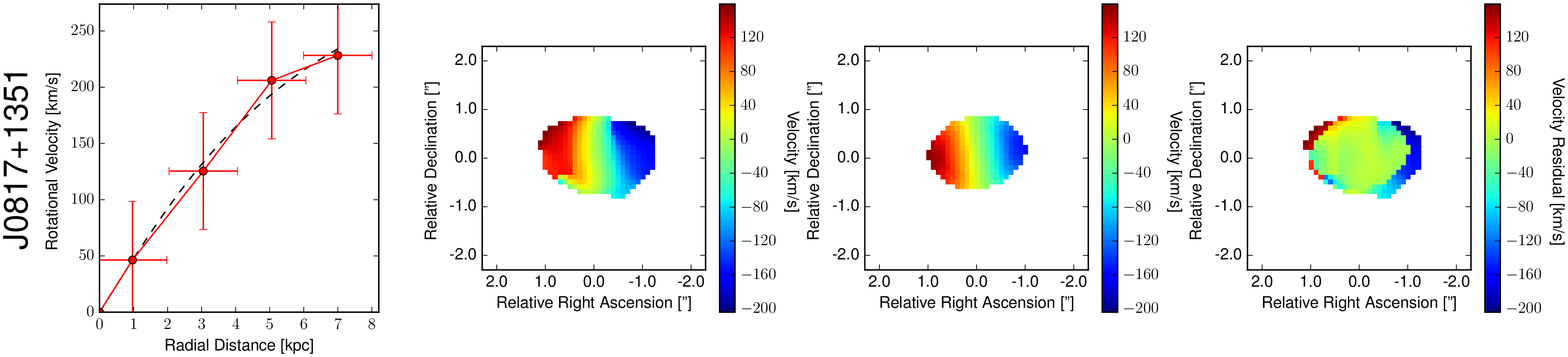}
\includegraphics[scale=0.4,clip=true]{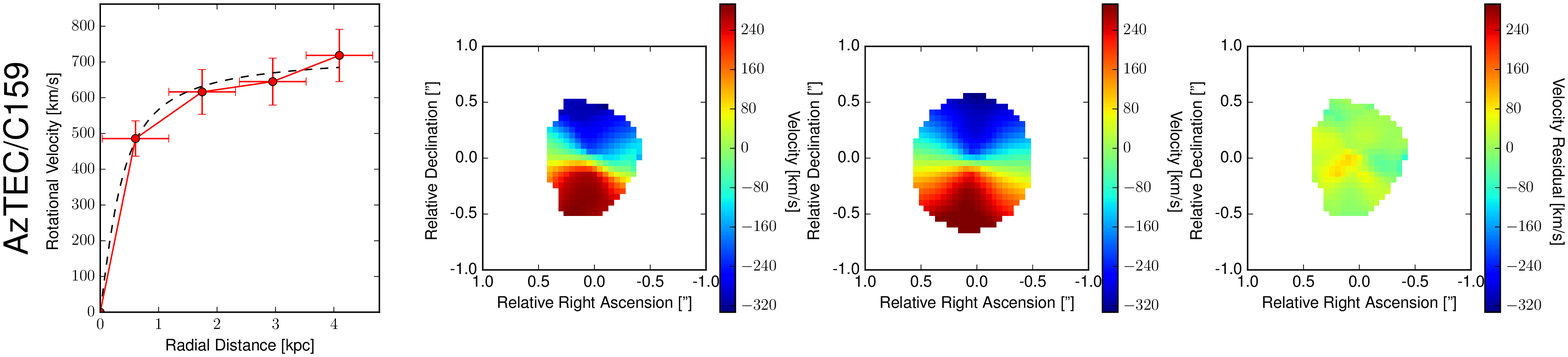}
\includegraphics[scale=0.4,clip=true]{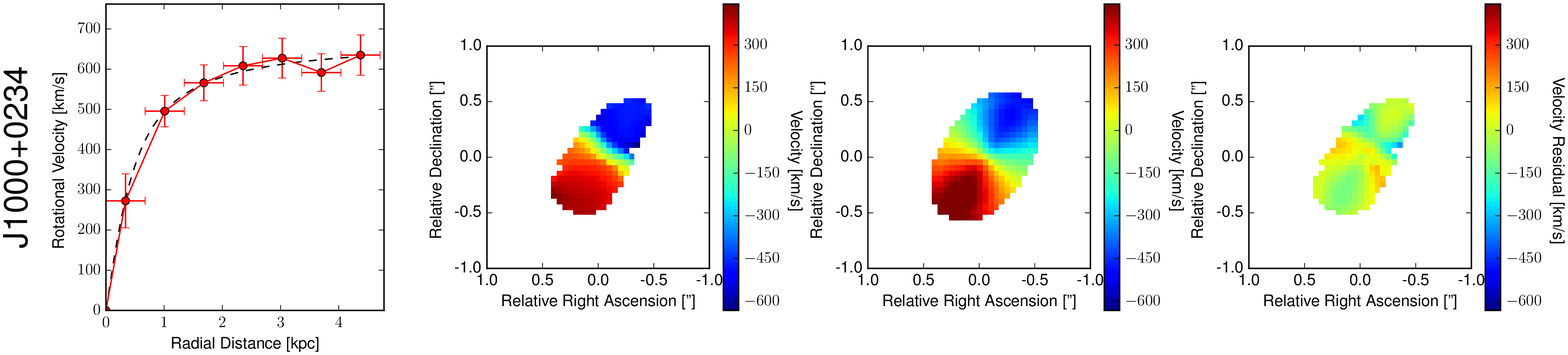}
\includegraphics[scale=0.4,clip=true]{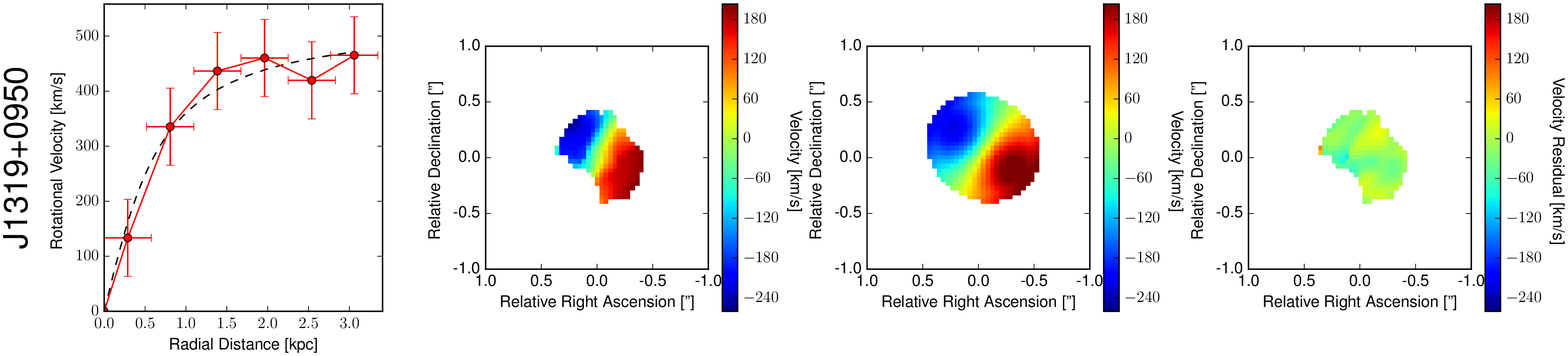}
\caption{Resulting images. \textit{Column 1:} rotation curve. \textit{Column 2:} data velocity field. \textit{Column 3:} fit model velocity field. \textit{Column 4:} residual (data - model) velocity field. Best fit systemic velocity (\textit{i.e.}, RC2) is subtracted from each velocity field.}
\label{velset}
\end{figure}

Figure \ref{velset} shows the rotation curves derived with \textlcsc{rotcur}, velocity fields created from the data, and the model velocity fields using the RC2 results. For the best-fit rotation curves, the vertical error bars are the uncertainty in $v_c$ (see Section \ref{RF}). The horizontal error bars represent the width of each ring. The red lines connecting each point are simply interpolations, while the dashed black lines are the better of a linear fit or a fit of the form $A\times\atan(x/B)$.


\subsection{Dynamical Mass}
\begin{figure}[b]
\centering
\includegraphics[scale=0.7,clip=true]{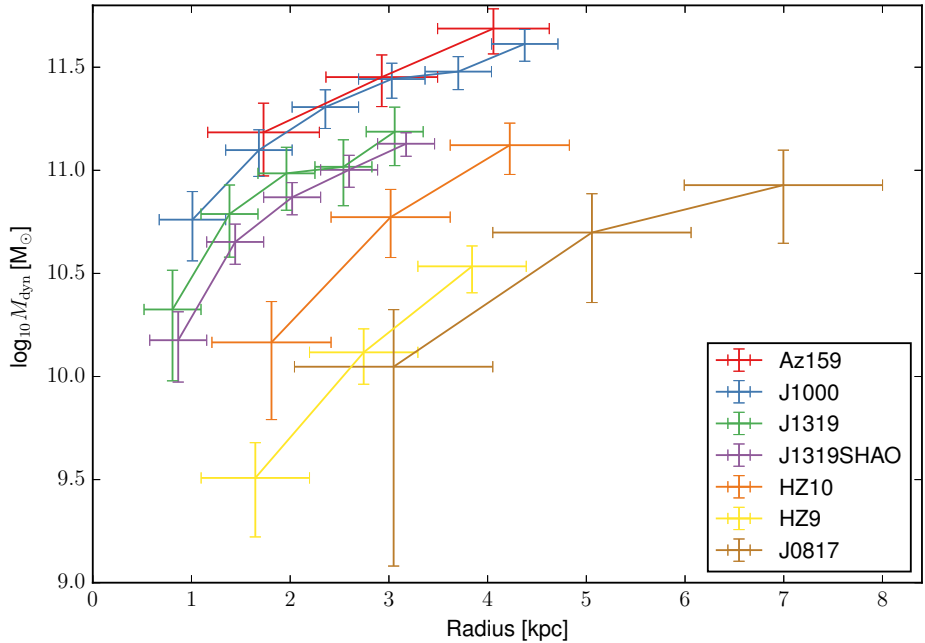}
\caption{Dynamical mass profiles for each galaxy. Due to resolution and sensitivity effects, a systematic uncertainty of $\sim1$\,dex also affects all points.}
\label{MP}
\end{figure}

An estimate of each source's dynamical mass could be found by assuming circular orbits, as we have done in the above analysis. This requires setting the centrifugal force equal to the gravitational attraction:
\begin{equation}\label{circle}
M_{\rm dyn}=\frac{v^2r}{G}
\end{equation} 
where $v$ is the velocity at radius $r$ and $G$ is the gravitational constant. Using the radius and velocity of the largest ring of our models results in the values shown in Table \ref{fitres2}. Equation \ref{circle} may also be applied to each ring individually, resulting in the mass profiles shown in Figure \ref{MP}.

In addition to our derived dynamical masses, we include previously found dynamical masses, dust masses, stellar masses, and H$_2$ masses. The previous dynamical masses were found using equation \ref{fwhm}, where the size and inclination angle was estimated from Gaussian fits to the integrated emission, assuming intrinsically circular disks. 

\pagebreak

\begin{deluxetable}{lccccc}
\tablecolumns{6}
\tablewidth{0pt}
\tablecaption{Galaxy Masses \label{fitres2}}
\tablehead{ \colhead{Source} & \colhead{log$_{10}M_{\rm dyn}$} & \colhead{log$_{10}M_{\rm dyn,lit}$} & \colhead{log$_{10}M_{\rm dust}$} & \colhead{log$_{10}M_{*}$} & \colhead{log$_{10}M_{\rm H_2}$}}
\startdata
HZ9             & $10.5\pm0.1$ 		   & $10.7^{+1.3\,a}_{-1.1}$ & --   		      & $9.9\pm0.2^{\,a}$  & -- \\ 
HZ10            & $11.1\pm0.1$            & $10.5^{+1.4\,a}_{-1.2}$ & --   		      & $10.4\pm0.2^{\,a}$ & --\\
J0817+1351 		& $10.9^{+0.2}_{-0.3}$    & $>10.8^{\,b}$ 		      & --	  		      & --				   & -- \\
AzTEC/C159      & $11.7\pm0.1$ 		   & -- 					  & $9.3\pm0.4^{\,c}$ & $11.0^{\,c}$	   & $10.0^{+0.2\,d}_{-0.1}$\\
J1000+0234      & $11.6\pm0.1$ 		   & $11.0^{\,e}$		      & --                & $8-10^{\,f}$       &  $10.4^{\,e}$\\
J1319+0950 	    & $11.2^{+0.1}_{-0.2}$ & $11.1_{-0.2}^{+0.1\,g}$ & $8.8^{\,h}$        & -- 				& $10.3\pm0.1^{\,h}$
\enddata
\tablecomments{References: 
\textit{a}-\cite{capa15}, 
\textit{b}-\citet{neel17},
\textit{c}-\citet{smol15},
\textit{d}-E. Jim\'{e}nez-Andrade et al. (in preparation), 
\textit{e}-\citet{schi08},
\textit{f}-\citet{capa08},
\textit{g}-\citet{wang13}, 
\textit{h}-\citet{wang11}.
All other values are from this paper.
}
\end{deluxetable}

We note that the uncertainties in the above masses are influenced only by the width of the ring and uncertainty in the velocity of the ring. While the velocity uncertainty does include the spectral resolution and \textlcsc{rotcur} uncertainty in $v_c$ \& $i$, there are still systematic uncertainties that we cannot estimate or account for. From an exploration of a model observation at different resolutions and S/N (see Appendix), we estimate a systematic uncertainty of $\sim1$ dex, to be added to the uncertainties shown in Figure \ref{MP}. When this uncertainty is included, our $M_{\rm dyn}$ ranges are similar to those already published. While our technique did not result in more precise measurements of the dynamical mass, it should be noted that our procedure is distinct from that used to find masses for these sources previously.

\subsection{HZ9}
While the velocity gradient observed in the data velocity field is replicated in the model, the two velocity fields show different East-West extents. In particular, while the other sources in our sample show elongation along the axis perpendicular to their isovelocity lines, HZ9 is nearly circular. This difference may imply that HZ9 is a nearly face-on disk that is rotating at a considerable rate, or that the distribution of emission is affected by our low resolution. The restoring beam was $0.67''\times0.53''$, resulting in square cells of $0.125''$. As seen in Figure \ref{velset}, the source only extends $\sim0.5''$ from its center, so few cells were used as input for each ring. To correct for this effect, models must be convolved with the observing beam and compared to the data on a channel-by-channel basis (see Section \ref{method2} for implementation of this method).

The difference in spatial extents between the model and data velocity fields is to blame for the disparity between the position angles derived by \textlcsc{rotcur} and zeroth moment imaging ($\sim7\sigma$ deviant). No combination of assumptions on other variables or variations of ring numbers produced a reliable rotation curve using the $\phi$ derived from Gaussian fitting, implying that it does not relate to the rotation. That the inclination angle estimates are similar (within $\sim2\sigma$) may act as marginal evidence that the matter is well organized into a rotating disk. This is supported by the agreement between the estimates of the center of the galaxy and the linear rotation curve.

\subsection{HZ10}
Similarly to HZ9, the linear rotation curve suggests solid rotation. However, the gradient in the data is not fully reproduced by the model fit. In particular, the southwest region of the data shows a high-velocity area that is not aligned with the eastern velocity gradient. 

Observations of \citet{pave16} show that the centers of \cii and [\ion{N}{2}] emission are offset both spatially and in velocity. One possible interpretation of this is an active merger, which we lack the resolution to account for here. This would partly explain why the best fit to the velocity field is dominated by the western velocity gradient of the galaxy, as the east could be galaxy interacting with the main galaxy.

While the effects of observing beam shape are less severe than in HZ9, the model velocity field again does not trace the integrated emission well. However, the inclination angles, positions, and redshifts agree between the \textlcsc{rotcur} and zeroth moment results.

\subsection{J0817+1351}
There is slight evidence for a flat rotation curve, and the model fits the general trend of the velocity field well. While a standard rotating galaxy shows a `butterfly'-like velocity field, with isovelocity lines radiating from its center, J0817+1351 simply shows a velocity gradient. The possible ripple in this field could be caused by a warp, but we lack the resolution to confirm this. While the systemic velocity, position angle, and inclination found through RC1 are more applicable than the GAUS estimates, the position from GAUS (only $\sqrt{2}$ pixels different) returns a better fit. 

The slight bend of the rotation curve at large radii may be interpreted as weak evidence for detection of disk-like rotation, but is more likely due to the non-uniform appearance of the northern and southern portions of the input velocity field.

\subsection{AzTEC/C159}
AzTEC/C159 shows the most likely evidence for a flat rotation curve at large radius. The use of a zeroth moment image here is not very helpful, as the integrated emission is nearly circular ($0.27''\times0.23''$). This introduces a large uncertainty in both inclination angle and position angle. However, all derived quantities agree with those found with \textlcsc{rotcur}. Through the variable combination analysis, the position angle found through the zeroth moment map was found to be inapplicable to the velocity field.

Since we have estimates of the stellar, dust, and molecular gas masses (Table \ref{fitres2} and references therein), we may estimate the total dark matter content ($M_{\rm DM}$) of the inner $r\sim4\,$kpc by assuming
\begin{equation}
M_{\rm dyn}=M_{\rm dust}+M_{*}+M_{\rm H_2}+M_{\rm DM}+\cdots
\end{equation}
where we are neglecting the mass of the ionized medium ($M_{\rm HII}$) and cold medium ($M_{\rm HI}$), for which we have no estimates. By inverting this equation, we find that $M_{\rm DM}\sim10^{11}$\,M$_{\odot}$. The implied baryon fraction ratio (M$_{\rm bary}$/M$_{\rm dyn}\sim0.3$) falls within the low end of the scatter of values found for local spiral galaxies by \citet{rich16}. 

\subsection{J1000+0234}
The velocity field of J1000+0234 is unique in this sample in that it features parallel isovelocity lines near its center and two spatially extended regions of high velocity gas at its far edges. That the central lines are parallel suggests that the gas velocity near the center rises linearly with increasing radius, as seen in Figure \ref{velset}. The best fit model to this velocity field is a steep linear rise near the galaxy center, followed by a flattened rotation profile. The flat portion of the curve produces the ``spider diagram''-like radiating isovelocity lines in the lobes, which are not seen in the data. 

The spatial distribution of emission and CO excitation level of this source has been interpreted as evidence for it being two galaxies in the midst of merging (\citealt{capa08,schi08}). While this would partially explain the appearance of the velocity field (namely, the lack of radiating isovelocity lines and the small velocity gradient across the northern and southern lobes), the apparent uniformity of the central gradient is evidence for this system being a single rotating galaxy. 

\subsection{J1319+0950}
Similarly to J1000+0234, J1319+0950 shows a local velocity decrease at the penultimate point in its rotation curve. Due to the non-elliptical shape of each galaxy, the outer rings are fit only to a few data points, instead of the full annuli available to the inner rings. However, as seen in Figure \ref{MP}, these local velocity minima do not translate to unphysical local dynamical mass minima.

\citet{shao17} fit a tilted ring model to this galaxy in a similar fashion, resulting in nearly identical parameters. In addition, they found a dynamical mass of log$_{10}M_{dyn}=11.1\pm0.1$, which agrees with the mass found through our analysis. The effects of slightly differing inclination angles ($29^{\circ}$ vs $33^{\circ}$) can be seen in the vertical offset in dynamical mass points in Figure \ref{MP}.

\section{\textlcsc{galmod} Modeling} \label{method2}
Our above approach has several drawbacks. One is that we cannot reliably account for the velocity dispersion ($\sigma_{\rm v}$) of these objects. \textlcsc{rotcur} does not allow the user to provide a prior estimate, fix the value, or include its effects in the model. Another limiting property is that we cannot control the density profile of the model. It is possible to estimate a surface density profile from the mass profile. However, this cannot be transformed into a physical mass density without the additional assumption of a vertical mass structure.

Another drawback is that we do not account for beam smearing. This effect, which is strongest for low resolution observations, acts to make strong velocity gradients appear less severe \citep{bege87,debl97,obri10,kamp15}. As shown in Figure 2 of \citet{debl97}, beam smearing alters the nature of the inner rotation curve but leaves the outer flat section relatively unchanged. Thus, while the underlying rotation may have a steeper initial rise, our detection of a signature of dark matter is still valid.

All of these drawbacks, in addition to possible deconvolution effects, can be accounted for by using the GIPSY task \textlcsc{galmod}. This takes in a radial mass density profile, scale height (z$_{\rm o}$), rotation curve, velocity dispersion, position angle, and inclination angle, outputting a model cube that matches the geometry (\textit{i.e.}, cell size, pointing center, channel width) of an input data cube. If the data cube has arbitrarily high resolution, the model cube may then be convolved with the restoring beam of an observation (\textit{e.g.}, \citealt{schu96}), and the observed and model cubes may be compared on a channel-by-channel basis, via a PV-diagram, or a global profile. If a sequence of such model cubes were created, spanning a range of density profiles, etc., they could be used to constrain the properties of the observed data. Alternatively, the Tilted Ring Fitting Code (TiRiFiC; \citealt{jozs07}) could be used to fit a model to the cube automatically. In addition to direct comparisons of spectral profiles, the cubes created by \textlcsc{galmod} may be mock observed using the CASA task \textlcsc{simobserve}. These observations may be used to test the validity of our spectral profiles and rotation curves.

It should be noted that \textlcsc{rotcur} has the advantages of speed and simplicity. However, these come at the price of forced assumptions and limited fitting ability. The use of \textlcsc{galmod} allows for the simultaneous exploration of a large number of variables (\textit{e.g.}, \citealt{yim14}) and can be used to explore the effects of beam smearing (\textit{e.g.}, \citealt{fran16,cald13}), but requires manual construction of a set of data cubes with different parameters, rather than automatically fitting for them. While TiRiFiC includes a data cube-based fitting routine, initial applications to our low S/N data resulted in poor fits. Thus, using \textlcsc{rotcur} to find the basic parameters of a source and then \textlcsc{galmod} to fit for other parameters while accounting for resolution effects seems to be the ideal path for exploring our datasets.

\subsection{Implementation}
To explore this avenue, we created a sequence of model cubes based on our data of AzTEC/C159 and J1319+0950, which were best fit by the \textlcsc{rotcur} process. \textlcsc{galmod} allows for the specification of a large number of parameters, so we adopted the rotation curves, position angles, central positions, and inclination angles found through our prior analysis to reduce any redundancies in our fitting process. Initial testing showed that results depended weakly on the input scale height and density profile, so these were assumed to be thin ($0.2\times$maximum radius; \citealt{lang17}) and flat ($\rho(r)=A$). Using these parameters, we created cubes with different velocity dispersions ($\sigma_{\rm v}=5$ to 145\,km\,s$^{-1}$). 

The CASA task \textlcsc{simobserve} was then used with these \textlcsc{galmod} cubes as input to simulate the ALMA cycle 1 observations of A. Karim et al. (in preparation) of AzTEC/C159 and the ALMA cycle 1 observations of \citet{shao17} of J1319+0950. It should be noted that the simulated and actual observations had slightly different beam sizes and channel sizes, but are still comparable.

In this way, we created a series of noiseless cubes and generated spectra of each, integrating over the entire galaxy. By comparing the spectra on a channel-by-channel basis with the spectrum of our original data cube, we calculated the goodness of fit.

\subsection{Results}
\begin{figure}
\centering
\includegraphics[scale=0.4]{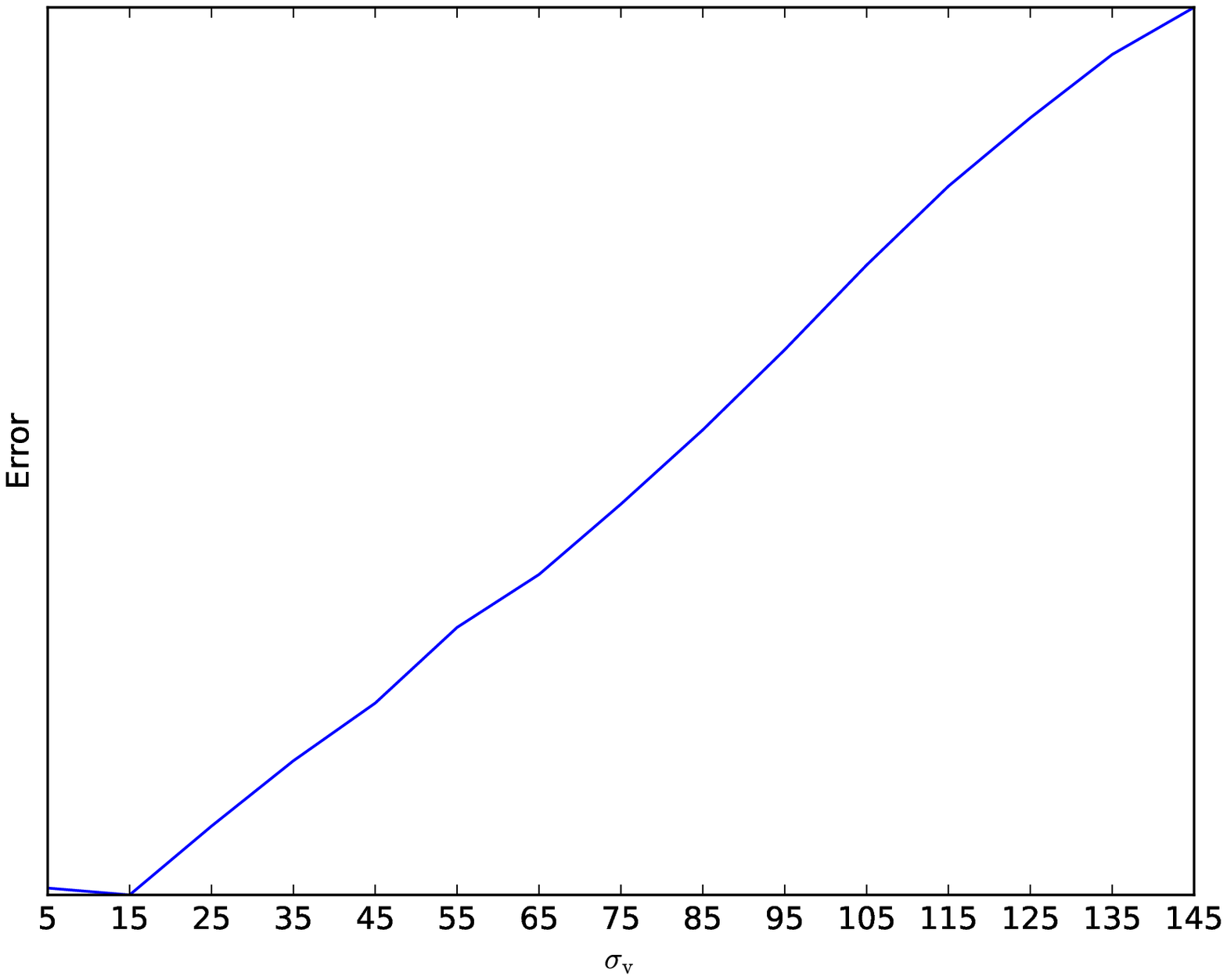}
\includegraphics[scale=0.4]{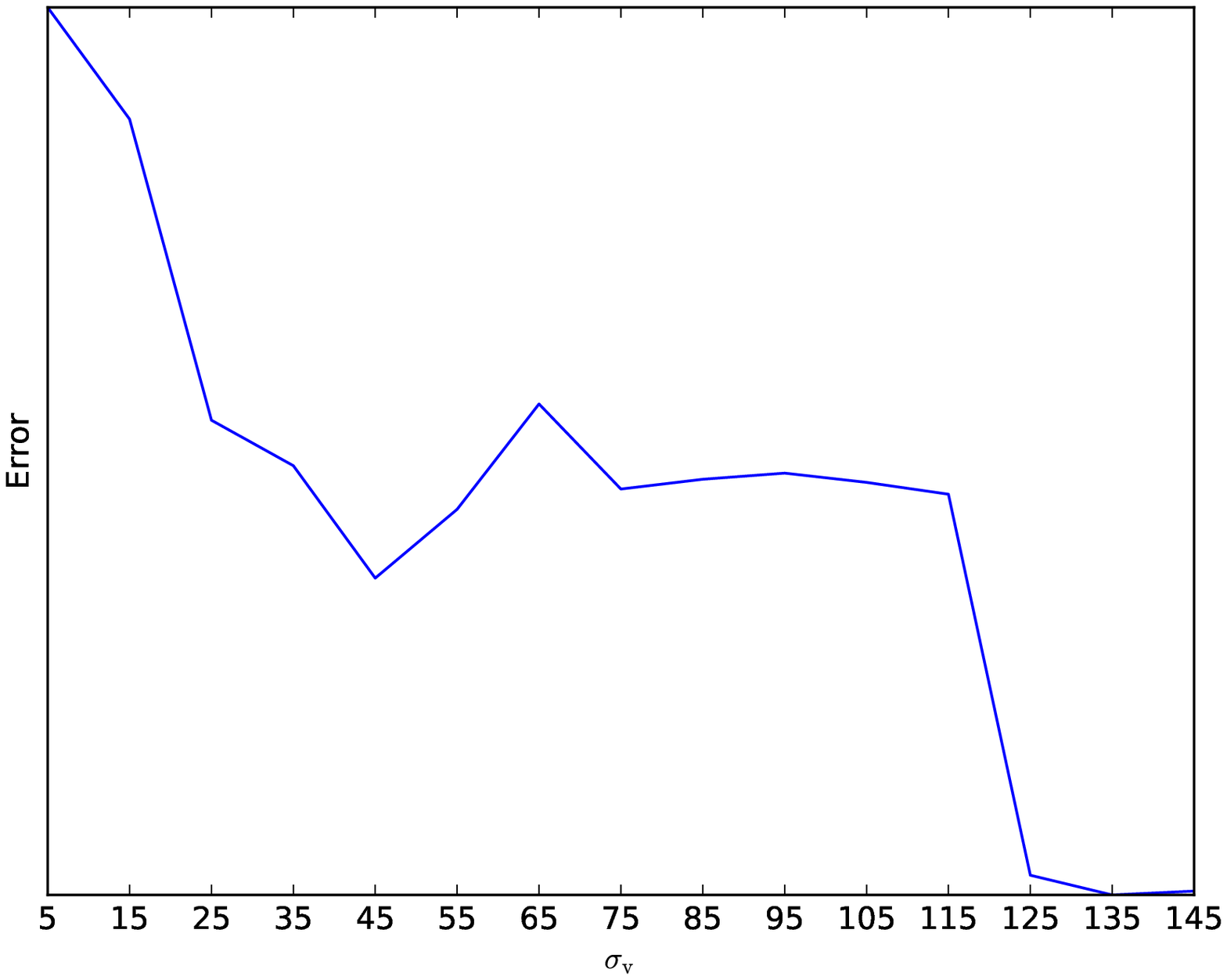}
\includegraphics[scale=0.4]{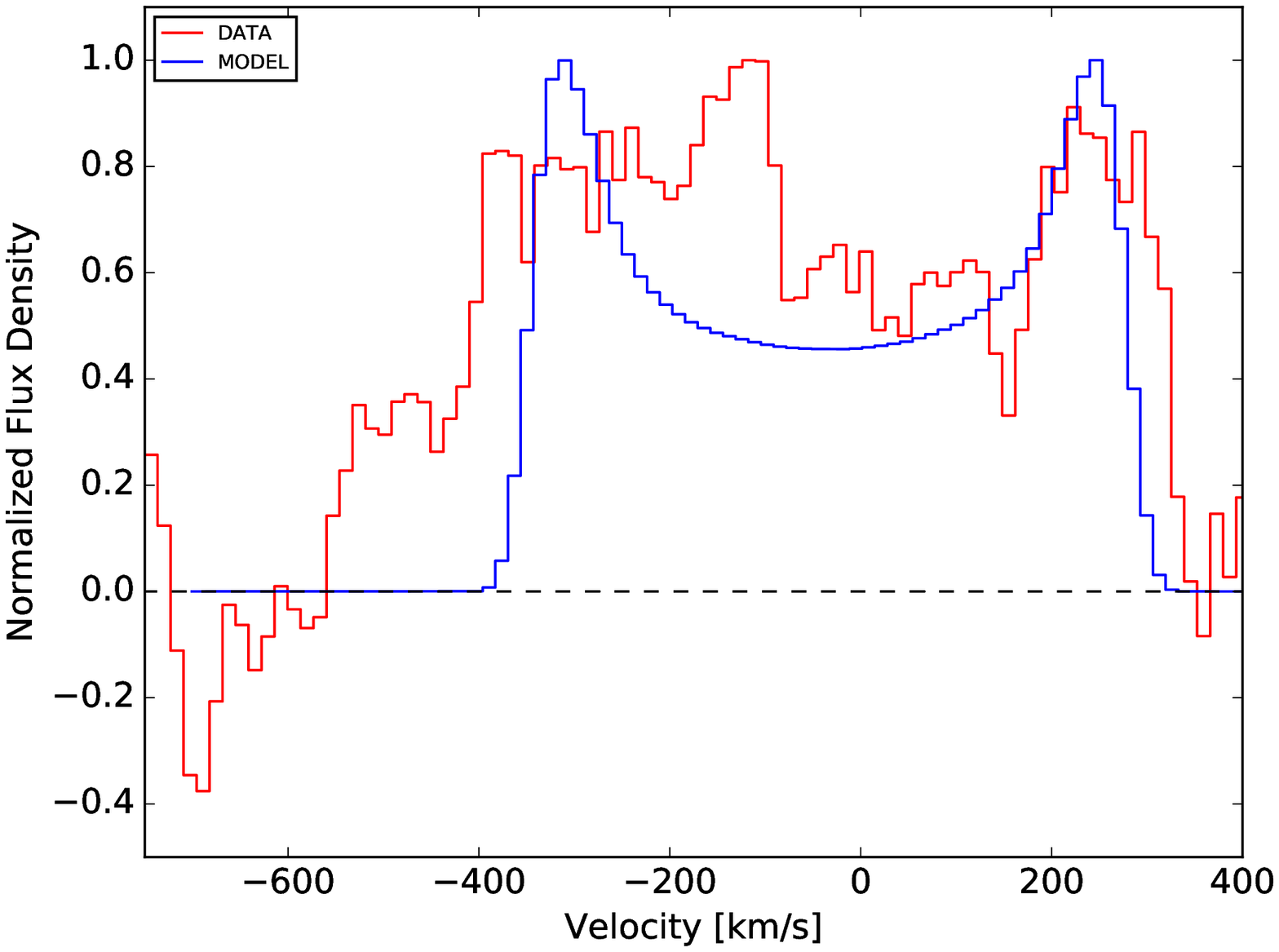}
\includegraphics[scale=0.4]{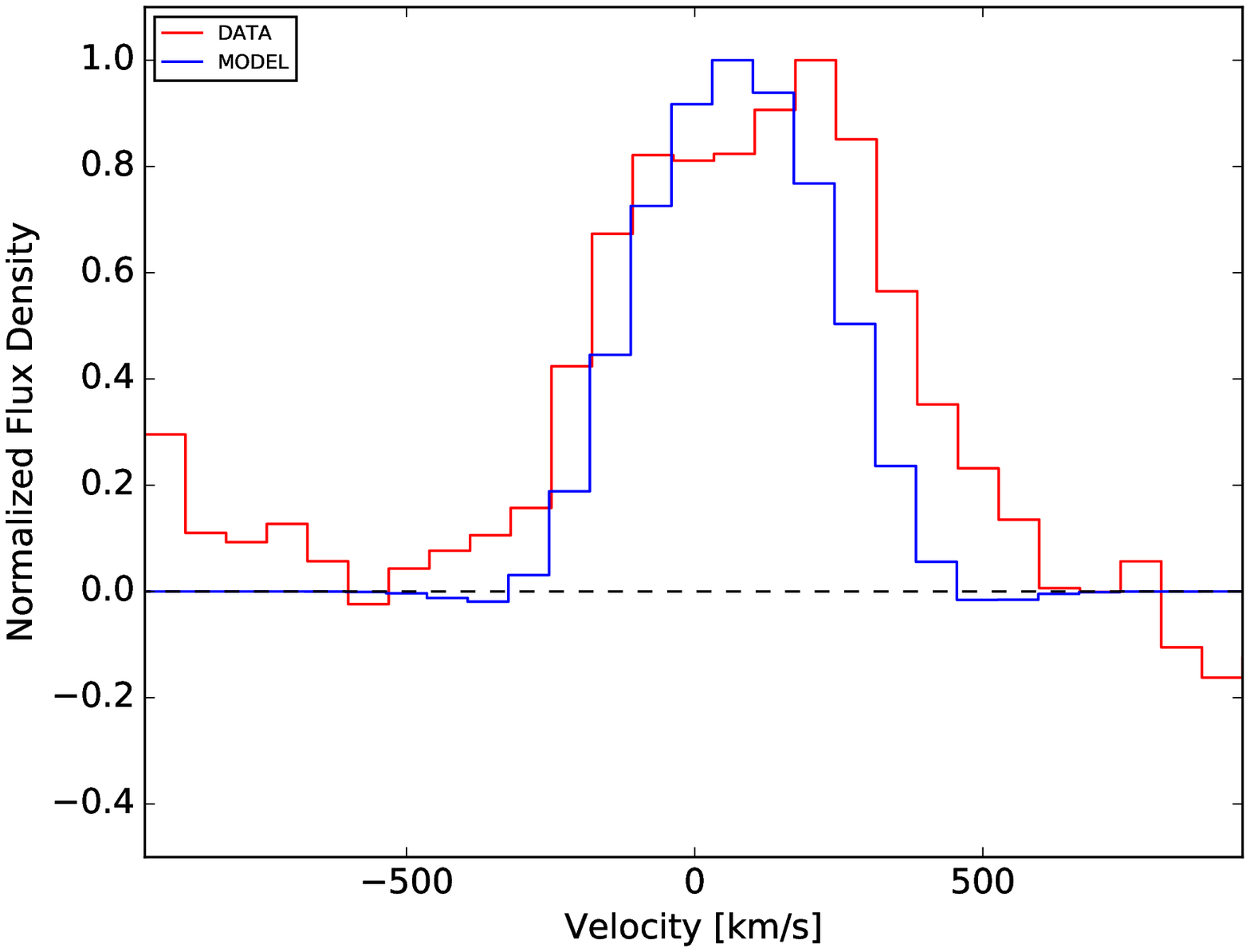}
\caption{TOP: Goodness of fit as a function of velocity dispersion for cubes created in \textlcsc{galmod} and \textlcsc{simobserve}.
BOTTOM: Comparisons of normalized spectra from data and best-fit model.
 In both rows, results for AzTEC/C159 are on the left, while those of J1319+0950 are on the right.}
\label{galmod}
\end{figure}

The resulting goodness of fit for each source as a function of velocity dispersion is shown in Figure \ref{galmod}.

For AzTEC/C159, models with small velocity dispersions were obviously favored. We adopt 15\,km\,s$^{-1}$ (which is the only minimum) as the velocity dispersion and compare the spectra of the model and data. While the model traces the general profile of the data, multiple features (\textit{e.g.}, the $\sim-500$\,km\,s$^{-1}$ and $\sim-180$\,km\,s$^{-1}$ components) are not replicated. The $-500$\,km\,s$^{-1}$ component is likely noise, as no significant emission is detected in its corresponding channels in the data cube. Instead, multiple low-level, diffuse noise sources are present.

In the case of J1319+0950, the dependence of the fitting error on velocity dispersion shows a significantly different behavior, with a global minimum at 135\,km\,s$^{-1}$. The width of the model spectrum is comparable to that of the data, although the model does not replicate the observed asymmetric horn profile, which may only be noise. 

We note that a full Bayesian approach (\textit{e.g.}, MultiNest; \citealt{fero08}) is preferable to our discrete exploration of the parameter space, but it is too computationally expensive at this time, given the low S/N of our data (\textit{i.e.}, runs take on the order of a day).

\section{Discussion} \label{PEA}

\subsection{Accuracy of Fits}
Using the method of tilted ring fitting, we are able to provide estimates for the position, position angle, inclination angle, and systemic velocity/redshift of a galaxy as a whole, in addition to its circular rotation velocity as a function of radius. If these parameters are used in the GIPSY task \textlcsc{galmod}, then we may also constrain the velocity dispersion, scale height, and density profile of the disk. However, since we are using these methods to fit relatively low S/N sources (\textit{i.e.}, with respect to observations of local galaxies that the methods were created to fit), some variables are not precisely constrained. While we will present a detailed analysis of systematic uncertainties and limitations in the process in a future paper (G. Jones, Y. Shao, et al., in preparation), we present our initial analysis in the Appendix.

To summarize, we are confident in our ability to determine the redshift, central position, and position angle of each source. However, the ambiguity between the rotational velocity and inclination of a disk makes the simultaneous determination of both quantities difficult. If the inclination angle could be found using a different technique (\textit{e.g.}, high resolution optical imaging, which is difficult for high-$z$, dusty galaxies), then this ambiguity could be removed, resulting in higher quality dynamical fits. In addition, the quality of the data (\textit{i.e.}, S/N, resolution, channel width) appears to be of less importance than the existence of an intrinsic rotating disk for this work. That is, while very poor quality data will return poor fits, both medium and high quality data will be nearly equally successful.

\subsection{Comparison With Other Techniques}
The dynamical characterization of $z>4$ galaxies with ALMA and \cii
is well complemented by recent work with IFU observations of H-$\alpha$
emissions of $z=0.6-2.6$ galaxies (\citealt{lang17,genz17}). The large sample size of 
Lang et al. allows for stacking of rotation curves, revealing a Keplerian-like 
decline of rotation speed at larger radii. Similar curves are detected for each of the bright sources
in Genzel et al. Comparison with dark matter 
models suggests a high baryon fraction and pressure support in these outer disks.
Since the outer portion of the rotation curves in this study are only tentatively detected,
we cannot confirm this.

In addition, rotation curves have already been found for ALMA \cii observations of $z>4$ galaxies, using other techniques. \citet{debr14} fit the velocity field of ALESS 73.1, a $z=4.76$ SMG, using a Bayesian implementation of the KINematic Molecular Simulation (KinMS) of \citet{davi13}, a simple arctan rotation curve model, and a dynamical model that accounts for beam smearing and assumes an exponential density profile. All three models returned good fits, resulting in a rotation curve and dynamical mass of $(3\pm2)\times10^{10}$\,M$_{\odot}$. Since KinMS creates data cubes for mock observations, with given surface brightness, rotation curve, scale height, velocity dispersion, pixel size, channel width, \textit{etc.}, it acts similarly to \textlcsc{galmod}, but is implemented in IDL.

Even if observations of high redshift sources are not well fit by tilted ring models or general dynamical codes, their dynamical masses may still be estimated using forms of equation \ref{fwhm}. In addition to some of the sources in our sample, this widespread method has been applied to the $z\sim4.1$ SMGs GN20.2a \& GN20.2b \citep{hodg13}, the $z\sim4.7$ SMG of BRI1202-0725 \citep{cari13}, the $z\sim6$ LBGs CLM1 \& WMH5 \citep{will15}, \textit{etc.}.

Thus, while our technique is not without precedent in the $z>4$ universe, it acts as a suitable complement to those procedures already in place.

\section{Conclusion} \label{conclude}

We have fit tilted ring dynamical models to \cii 158$\mu$m line
observations of two MS galaxies (HZ9 \& HZ10), a QSO host galaxy (ULAS J1319+0950), 
two SMGs (AzTEC/C159 \& COSMOS J1000+0234), and a DLA host galaxy (ALMA J0817+1351), all at $z>4$. 
Our dynamical analysis is one of the the first
attempts to go beyond very simple conclusions based on fitting
integrated line profiles for $z > 4$ galaxies, using the spatial
information inherent in the interferometric observations. The three
low-luminosity galaxies (HZ9, HZ10, and ALMA J0817+1351) show linear 
velocity gradients, consistent with
either rotation or other dynamical models, and limited by
signal-to-noise and resolution. The three luminous galaxies (ULAS J1319+0950, AzTEC/C159, and J1000+0234) 
exhibit possible evidence for flattening of the velocity field at large radii,
more suggestive of rotating disk galaxies. In the case of AzTEC/C159,
evidence for $\mathcal{O}(10^{11})$\,M$_{\odot}$ of dark matter was found.

In addition to fitting models to velocity fields, we created a sequence
of model data cubes spanning a range of possible velocity dispersions
for AzTEC/C159 and J1319+0950. These high resolution
model cubes were then observed using the CASA task
\textlcsc{simobserve}, and then compared to the integrated
line profile.By comparing our data to these models, we found evidence
for low ($\sim15$\,km,s$^{-1}$) and moderate ($\sim85$\,km\,s$^{-1}$) velocity dispersions. 

In an effort to quantify the accuracy of our technique, we tested our code on mock data cubes at two levels of resolution and sensitivity. From this, we learned that high sensitivity and resolution result in better fits, but the effects of beam smearing are still evident. Because of the above uncertainties, our conclusions are broad. For each of the three MS galaxies (HZ9, HZ10, and J0817+1351), the mainly linear rotation curves are likely non-physical and actually caused by beam smearing effects. On the other hand, the two SMGs and the QSO host galaxy in this sample (AzTEC/C159, J1000+0234, J1319+0950), show stronger evidence for a flattened rotation curve. A true interpretation of these results must wait until higher resolution observations are performed, allowing us to truly probe the dynamical nature of these sources. However, these initial results are promising.

While full dynamical characterization or rotation curve decomposition
are not yet possible, these first results using ALMA data (taken
mainly with short integration times and relatively few antennas) are
encouraging. As ALMA attains full capability, high resolution, higher
signal-to-noise observations of the \cii 158$\mu$m line will allow 
for detailed determination of the galaxy dynamical masses at the highest
redshifts.  In parallel, large new facilities in the near-IR
will determine the stellar profile of early galaxies, while future facilities
such as the Next Generation Very Large Array, will determine gas density 
profiles at high resolution (\citealt{cari15,mcki16,seli17}). Together, these 
facilities will allow for a full analysis of the baryonic and dark matter
components of galaxies out to cosmic reionization. 

\acknowledgements
This work makes use of the following ALMA data: 2011.0.00206.S, 2012.1.00240.S, 2012.1.00523.S, 2012.1.00978.S, and 2015.1.01564.S. ALMA is a partnership of ESO (representing its member states), NSF (USA), and NINS (Japan), together with NRC (Canada), NSC and ASIAA (Taiwan), and KASI (Republic of Korea), in cooperation with the Republic of Chile. The Joint ALMA Observatory is operated by ESO, AUI/NRAO, and  NAOJ.  The  National  Radio Astronomy Observatory is a facility of the National Science Foundation operated under cooperative agreement by Associated Universities, Inc. GJ is grateful for support from NRAO through the Grote Reber Doctoral Fellowship Program. DR and RP acknowledge support from the National Science Foundation under grant number AST-1614213 to Cornell University. RP acknowledges support through award SOSPA3-008 from the NRAO. AK acknowledges support by the Collaborative Research Council 956, sub-project A1, funded by the Deutsche Forschungsgemeinschaft (DFG).


\pagebreak

\appendix 
In an effort to quantify the reliability of our fitting routine, we apply it to a series of simulated observations of a well-studied galaxy. The key to this analysis is that all six of these observations are based the same galaxy, with identical intrinsic dynamics and morphology. Thus, this exploration allows us to test the fitting routine without including scatter from differing source properties.

\section{Model Creation}
To construct our sky models, we use observations of M51 (NGC5194) in CO($J=1\rightarrow0$) emission taken by \citet{helf03} with the BIMA interferometer between 1997 and 1999. These data were transformed to represent emission from a $z=4.2$ galaxy with a similar velocity width and flux density as J0817$+$1351\footnote{See \url{https://casaguides.nrao.edu/index.php/Simulation_Recipes} for general procedure.}. In short: the frequency of the first channel was shifted to that of \cii at $z=4.2$, the width of each channel was multiplied by the approximate ratio of the J0817+1351 linewidth to the M51 linewidth (450/170), and the peak flux density was adjusted to match J0817+1351. This sky model was then used in the CASA task \textlcsc{simobserve} with different ALMA configurations and observation lengths.

The mock datasets are created using two different resolutions: $0.50''\times0.43''$ (low, L) and $0.30''\times0.27''$ (medium, M). Observations of each were simulated using both 3\,hours on source (H3) and 30 hours on source (H30). In total, we here consider four simulated observations, exploring two levels of sensitivity and two levels of resolution. Each mock observation is then folded through our analysis in a manner identical to the observed sources in the above paper.

This source presents a challenge, as it exhibits both variation in inclination angle ($i=+24^{\circ}$ to $-24^{\circ}$; \citealt{oika14}) and distinct spiral arms \citep{shet07}, both of which are unaccounted for in our code. However, this example is more useful than an idealized test (\textit{e.g.}, using a \textlcsc{galmod} dataset), as we are able to determine how these non-regular features affect our results at different resolutions and sensitivities. In addition, our models at most extend to $10$\,kpc in our adjusted frame, or $8.3$\,kpc in the frame of M51, which is an area of relatively small inclination variation (Figure 5 of \citealt{oika14}).

\section{Results}

\begin{figure}[b]
\centering
\includegraphics[scale=0.4,clip=true]{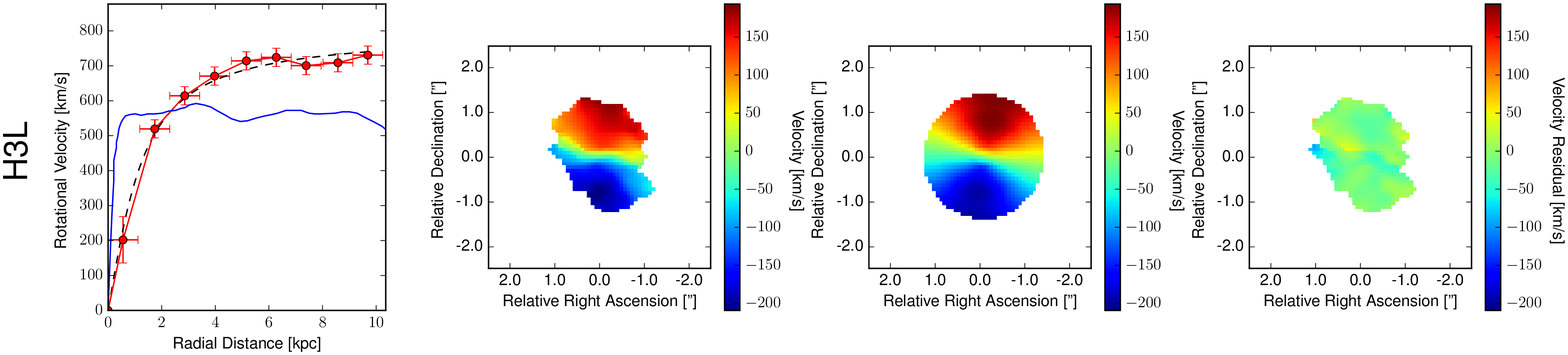}
\includegraphics[scale=0.4,clip=true]{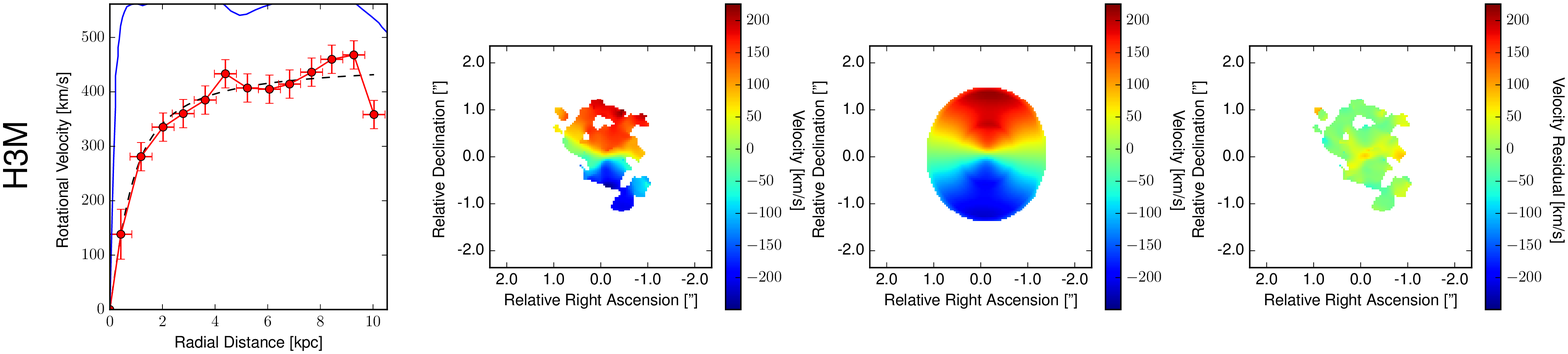}
\includegraphics[scale=0.4,clip=true]{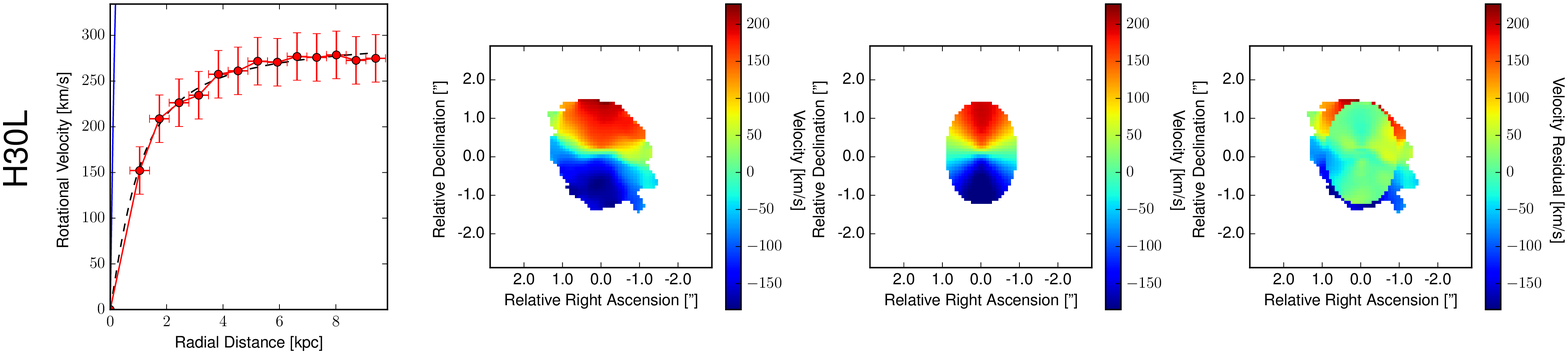}
\includegraphics[scale=0.4,clip=true]{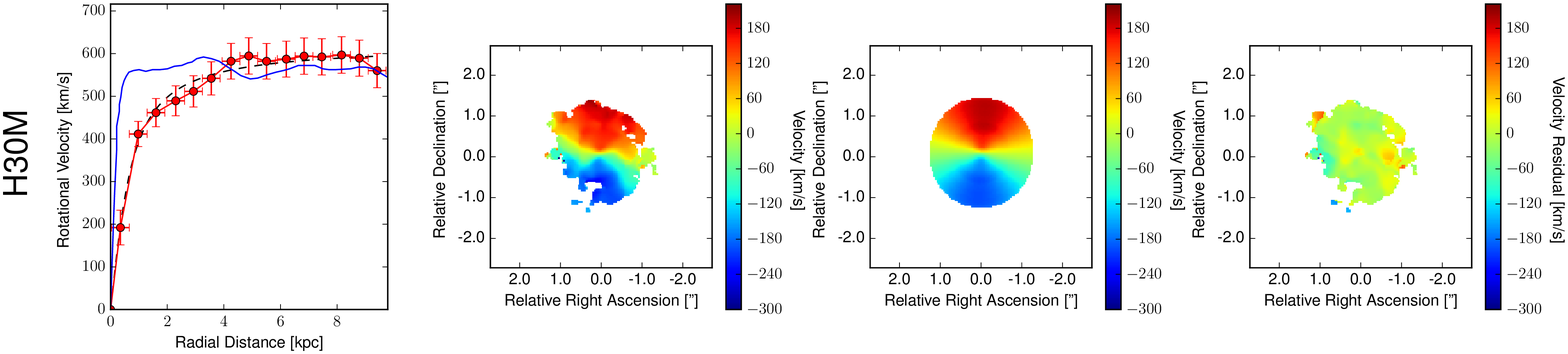}
\caption{Resulting images. \textit{Column 1:} rotation curve. \textit{Column 2:} data velocity field. \textit{Column 3:} fit model velocity field. \textit{Column 4:} residual (data - model) velocity field. Best fit systemic velocity (\textit{i.e.}, RC2) is subtracted from each velocity field.}
\label{velset_MO}
\end{figure}

\begin{figure}
\centering
\includegraphics[scale=0.7,clip=true]{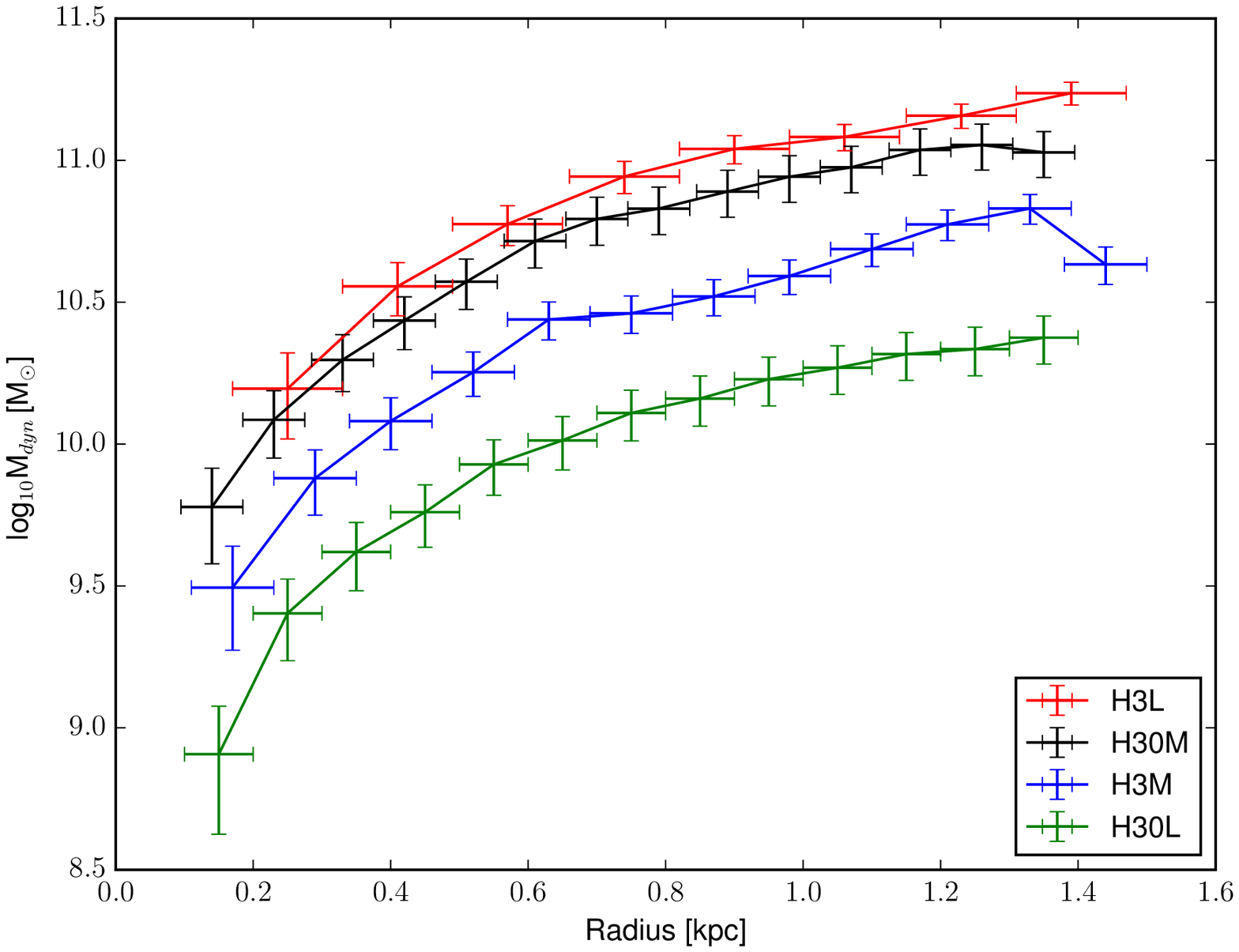}
\caption{Dynamical mass profiles for each galaxy.}
\label{MP_MO}
\end{figure}

Below are the best fit results from rotcur (Table \ref{approt}), the rotation curves, velocity fields, and residual maps (Figure \ref{velset_MO}), and a mass profile plot (Figure \ref{MP_MO}). 

\begin{deluxetable}{lc|cccccc}
\tablecolumns{8}
\tablewidth{0pt}
\tablecaption{Rotcur Fitting Results \label{approt}}
\tablehead{ \colhead{Source} & & \colhead{$z_{\rm fit}$} & \colhead{$\phi$ [$^{\circ}$]} & \colhead{$i$ [$^{\circ}$]} & \colhead{RA} & \colhead{Dec} & \colhead{log\,$M_{\rm dyn}$ [M$_{\odot}$]}}
\startdata
H3L  & RC1  & $4.1943(5)$ & $-15\pm1$   & $24\pm1$  & 13h29m52.39(2)s  & $-30^{\circ}00'0.0(4)''$&\\ 
     & GAUS & $4.1936(6)$ & $31\pm11$   & $46\pm7$  & 13h29m52.39(2)s  & $-30^{\circ}00'0.1(4)''$&\\ 
     & RC2  & $4.1941(5)$ & $-11\pm5$   & $16\pm1$  & 13h29m52.39(2)s  & $-30^{\circ}00'0.1(4)''$& $11.24\pm0.04$\\[2mm]
H3M  & RC1  & $4.1941(5)$ & $-4\pm4$   & $31\pm9$  & 13h29m52.34(2)s  & $-30^{\circ}00'0.0(2)''$&\\ 
     & GAUS & $4.1943(5)$ & $8\pm17$   & $42\pm11$  & 13h29m52.35(2)s  & $-30^{\circ}00'0.0(2)''$&\\ 
     & RC2  & $4.1941(5)$ & $-3\pm1$   & $28\pm3$  & 13h29m52.34(2)s  & $-30^{\circ}00'0.0(2)''$& $10.6\pm0.1$\\[2mm]
H30L & RC1  & $4.1937(5)$ & $1\pm1$   & $44\pm1$    & 13h29m52.35(3)s  & $-30^{\circ}00'0.0(4)''$&\\ 
     & GAUS & $4.1944(7)$ & $33\pm11$   & $40\pm9$  & 13h29m52.35(3)s  & $-30^{\circ}00'0.0(4)''$&\\ 
     & RC2  & $4.1938(5)$ & $-1\pm1$   & $47\pm1$   & 13h29m52.35(3)s  & $-30^{\circ}00'0.0(4)''$& $10.4\pm0.1$\\[2mm]
H30M & RC1  & $4.1942(5)$ & $-5\pm3$   & $18\pm4$   & 13h29m52.35(2)s  & $-30^{\circ}00'0.0(2)''$&\\ 
     & GAUS & $4.1945(6)$ & $27\pm12$   & $41\pm8$  & 13h29m52.35(2)s  & $-30^{\circ}00'0.0(2)''$&\\ 
     & RC2  & $4.1942(5)$ & $-7\pm4$   & $19\pm5$   & 13h29m52.35(2)s  & $-30^{\circ}00'0.0(2)''$& $11.0\pm0.1$
\enddata
\end{deluxetable}

\section{Discussion}

We now consider the effects of resolution and sensitivity on the fit parameters, rotation curves, and mass profiles.

All models yield similar redshifts, which are found by directly converting the best-fit systemic velocity. This parameter, which influences a velocity field as an overall shift, is therefore derivable using even poor data. Note that the first channel in the cube (\textit{i.e.}, crval3) was set as the rest frequency of \cii at $z=4.2$, so the redshifts here are all less than $4.2$. Similarly to the redshifts, the central position of each model galaxy agrees.

The position and inclination angles show more variability between datasets. Both high sensitivity position angle estimates agree with $\phi\sim0^{\circ}$ from \citet{rega01} to within $2\sigma$, while the two low sensitivity datasets are $2-3\sigma$ discrepant. Thus, our determination of position angle is weakly dependent on sensitivity, but even low sensitivity observations may yield semi-accurate position angles.

On the other hand, the inclination of each dataset varied from $16^{\circ}$ to $47^{\circ}$. \citet{rega01} state an inclination angle of $15^{\circ}$, while \citet{oika14} use $24^{\circ}$. Since an ambiguity exists between inclination angle and circular velocity (\textit{i.e.}, $V(x,y)\propto v_{\rm c}(r)\sin\,i$), this spread in inclination angles reflects a similar spread in rotation velocity. 

We have scaled the rotation curve of \citet{oika14} (their Figure 1) and plotted it over our mock rotation curves (Figure \ref{velset_MO}). In all cases, the effects of beam smearing are obvious. The inner, steep rise of the scaled M51 rotation curve is not traced, and the initial rise modeled rotation curve is less severe. The only curve that agrees is that of H30M. Discrepancies between the other curves are caused by the scatter in their inclination angles, especially the high sensitivity, low resolution case ($i>40^{\circ}$). This uncertainty in rotation curve scale extends to an uncertainty in dynamical mass.

If this method was unaffected by the sensitivity and resolution of a given observation, then the mass profiles of Figure \ref{MP_MO} would be coincident. Instead, the two low-resolution datasets are outliers, and the large-radius dependence of the other models is disparate. Interestingly, the two medium-resolution datasets are highly similar. However, H3M shows a decrease in dynamical mass with increased radius, which is unphysical and suggests that the model should only be trusted at the inner radii.

We have also applied the technique of \citet{shao17} to the six datasets, resulting in mainly the same best-fit parameters. The only parameters that differed between our two analyses were the inclination angles of the two low resolution datasets (\textit{i.e.}, H3L \& H30L). Namely, the procedure of Shao et al. finds $i=46\pm9^{\circ}$ for H3L (the above procedure yields $16\pm1^{\circ}$) and $24\pm5^{\circ}$ for H30L ($47\pm1^{\circ}$). However, since the low resolution datasets show strange mass profiles, this difference in inclination angles may only signify that the underlying datasets are difficult to model, but the two analysis methods are in agreement.

\section{S/N Effects}

The quality of each fit may dependent on multiple characteristics, including the spatial resolution of the disk, the S/N of the detection, the number of channels in which line emission is present, and whether or not the source is in reality an ordered, rotating disk. While we cannot control the last property, it is possible to compare the others. In an attempt to identify which observation characteristics contribute to a good fit, we tabulate the fractional resolution of the source ($R_{\rm max}/$HWHM of the minor axis of the restoring beam), the S/N of the \cii detection, and the fractional velocity resolution (\textit{i.e.}, the number of channels with line emission).

\begin{deluxetable}{lcccc}
\tablecolumns{5}
\tablewidth{0pt}
\tablecaption{Observation Characteristics \& Fit Quality \label{quality}}
\tablehead{ \colhead{Source} & \colhead{$R_{\rm max}/$HMWM$_{minor}$} & \colhead{S$/$N} & \colhead{v$_{\rm range}/\Delta$v$_{\rm channel}$} & \colhead{Fit Quality}}
\startdata
HZ9             & 2.6 & 9.5  & 19.4 & 4 \\ 
HZ10            & 3.5 & 15.5 & 16.4 & 3 \\
J0817+1351 		& 2.7 & 10.7 & 7.0  & 2 \\
AzTEC/C159      & 3.0 & 11.2 & 44.7 & 1 \\
J1000+0234      & 4.8 & 16.0 & 43.1 & 2 \\
J1319+0950 	    & 5.5 & 6.8  & 6.6  & 2 \\ \hline
H3L              & 8.4 & 14.7 & 15.5 & 2 \\
H3M				 & 11.1 & 9.4 & 18.3 & 3 \\
H30L				& 6.5 & 23.5 & 15.9 & 4 \\
H30M				& 10.4 & 12.0 & 17.0 & \,\,1
\enddata
\end{deluxetable}

In Table \ref{quality}, we also estimate the `Fit Quality' of each source, where 1 represents the best fit and 4 the worst. Since known rotation curves for our other sources do not exist, their quality values are qualitatively based on the agreement of the model and data velocity fields. Az159 shows excellent agreement, while J0817+1351, J1000+0234, and J1319+0950 show slight deviations. Only portions of HZ9 and HZ10 are well fitted, but HZ9 is also poorly spatially sampled. The modeled sources were simply ordered by how well their rotation curves agreed with that of \citet{oika14}.

The spatial and spectral resolution of each source do not show obvious trends with the fit quality. Curiously, both HZ9 and J0817+1351 show similar spatial fractional resolution, but since J0817+1351 has twice the number of cells per HMWM$_{minor}$ (8.8 vs 4.2 for HZ9), it appears better resolved. From this analysis, it would appear that poor fits will be returned if any one of these characteristics is weak (\textit{e.g.}, poor spatial fractional resolution of HZ9/H30L).

In summary, our analysis is best applied to medium resolution, high sensitivity data of orderly rotating disks. Intermediate resolution allows for characterization of the galaxy's rotation as a whole, while minimizing under-resolution effects. Higher sensitivity presents more of the outer disk, while improving the results in the easily detected inner disk. The central positions, position angles, and systemic velocities are accurately determined, even in the case of low resolution/sensitivity observations. The ambiguity between rotational velocity and inclination angle introduces an uncertainty in inclination and a scaling factor in the rotation curve and mass profile. This may be corrected for by determining the inclination angle separately (\textit{e.g.}, using axis ratios of resolved observations). 

\end{document}